\documentclass[aps,prb,10pt,reprint,showpacs,amssymb,amsmath,amsfonts,floatfix]{revtex4-1}
\usepackage{undertilde}
\usepackage[pdftex]{graphicx}

\def\El{\mathcal{L}}
\def\kv{\mathbf{k}}
\def\pv{\mathbf{p}}
\def\sv{\mathbf{s}}
\def\Sv{\mathbf{S}}
\def\qv{\mathbf{q}}
\def\sh{\hat{\mathbf{s}}}
\def\zh{\hat{\mathbf{z}}}
\def\kh{\hat{\mathbf{k}}}
\def\ph{\hat{\mathbf{p}}}
\def\th{\hat{\mathbf{t}}}
\def\nh{\hat{\mathbf{n}}}
\def\qh{\hat{\mathbf{q}}}
\def\Ov{\mathbf{\Omega}}
\def\vt{\vartheta}
\def\vp{\varphi}
\def\sigv{\boldsymbol{\sigma}}
\def\one{\hat{1}}
\def\ve{\varepsilon}
\def\Tr{\mathrm{Tr}}
\def\l{\left}
\def\r{\right}
\def\ua{\uparrow}
\def\da{\downarrow}
\def\qt{\widetilde{q}}
\def\ee{\textit{e--e }}
\def\eh{\textit{e--h }}
\def\ei{\textit{e--i }}

\begin{document}
\title{D'yakonov-Perel' spin relaxation for degenerate electrons in the electron-hole liquid}
\author{Matthew D. Mower}
\affiliation{Department of Physics and Astronomy, University of Missouri, Columbia, Missouri 65211, USA} 
\author{G. Vignale}
\affiliation{Department of Physics and Astronomy, University of Missouri, Columbia, Missouri 65211, USA}
\author{I. V. Tokatly}
\affiliation{Nano-Bio Spectroscopy group and ETSF Scientific Development Centre, 
Departamento de F\'isica de Materiales, Universidad del Pa\'is Vasco UPV/EHU, 
 E-20018 San Sebasti\'an, Spain}
\affiliation{IKERBASQUE, Basque Foundation for Science, E-48011 Bilbao, Spain}

\begin{abstract}
We present an analytical study of the D'yakonov-Perel' spin relaxation time for degenerate electrons in a photo-excited electron-hole liquid in intrinsic semiconductors exhibiting a spin-split band structure. The D'yakonov-Perel' spin relaxation of electrons in these materials is controlled by electron-hole scattering, with small corrections from electron-electron scattering and virtually none from electron-impurity scattering. We derive simple expressions (one-dimensional and two-dimensional integrals respectively) for the effective electron-hole and electron-electron scattering rates which enter the spin relaxation time calculation. The electron-hole scattering rate is found to be comparable to the scattering rates from impurities in the electron liquid -- a common model for $n$-type doped semiconductors. As the density of electron-hole pairs decreases (within the degenerate regime), a strong enhancement of the scattering rates and a corresponding slowing down of spin relaxation is predicted due to exchange and correlation effects in the electron-hole liquid. In the opposite limit of high density, the original D'yakonov-Perel' model fails due to decreasing scattering rates and is eventually superseded by free precession of individual quasiparticle spins. 
\end{abstract}


\maketitle

\section{Introduction}
Semiconductor spintronic devices are poised to complement existing electronic devices in the future by using the electron's spin degree of freedom to store and transfer information.\cite{np_3_153,science_294n5546_1488} Of the many branches and uses of spintronics, spin relaxation is relevant to each. The rate of decay of spin polarization needs to be taken into consideration in spin packet transportation, spin transistor junctions, spin qubits, and the like. Additionally, microscopic studies of spin relaxation severely test our ability quantitatively describe effective interactions and scattering mechanisms in the solid state.\cite{fabian_aps,rmp_76_323,opt_orient,Bronold200473,Wu201061}

The most complete study of spin relaxation in semiconductors to date is by Jiang \textit{et al.},\cite{prb_79_125206} who made calculations of the spin relaxation time (SRT) due to each of the relevant spin relaxation mechanisms in III-V semiconductors: Bir-Aronov-Pikus (BAP) \cite{spjetp_42_705}, D'yakonov-Perel' (DP)\cite{dyak_1}, and Elliott-Yafet (EY)\cite{pr_85_478,*pr_96_266}. Comparing the relative efficiencies of the mechanisms, their results suggest that D'yakonov-Perel' spin relaxation is dominant for essentially all the electron densities and temperatures of experimental interest.\cite{prb_79_125206} Although Teng \textit{et al.} initially found cases where BAP spin relaxation dominates DP spin relaxation in intrinsic GaAs,\cite{prb_66_035207} it was later pointed out by Jiang and Wu that non-degenerate statistics were being applied to degenerate electrons.\cite{jpd_42_238001} Also noted by Jiang \textit{et al.}, Song and Kim have investigated spin relaxation due to all of the relevant spin relaxation mechanisms in $n$- and $p$-type III-V semiconductors,\cite{prb_66_035207} but their analytical expressions are again only applicable to the non-degenerate regime. Tamborenea \textit{et al.}\cite{prb_68_245205} have calculated EY SRTs due to electron-impurity and electron-electron collisions in $n$-doped GaAs for a wide range of temperatures and electron densities, finding that the EY mechanism alone is insufficient to explain experimental SRT measurements,\cite{prl_80_4313} suggesting that DP spin relaxation may account for the discrepancies. These findings encourage us to continue looking at DP spin relaxation as the primary spin relaxation mechanism in GaAs and similar semiconductors of the zincblende structure.

The focus of the present study is on the role played by many-body interactions on the DP spin relaxation mechanism -- a role that we would like to clearly disentangle from that of other scattering mechanisms. The theory of many-body effects in DP spin relaxation for electrons in semiconductors was formulated by Glazov and Ivchenko (GI) several years ago.\cite{jetpl_75_8_403,jetp_99_6_1279} The basic idea is that electron-electron scattering causes the electron quasiparticle to perform a random walk in momentum space; this in turn causes random variations of the direction and magnitude of the spin precession axis. DP spin relaxation arises from the cumulative effect of many small precessions about randomly varying axes, and its main signature is that the spin relaxation rate is inversely proportional to the momentum scattering rate. In practice, however, it is very difficult to disentangle the contribution of the electron-electron scattering rate from the similar but much larger contribution of electron-impurity scattering, which is inevitably present in doped semiconductors. An important exception arises in intrinsic semiconductors, when a non-equilibrium population of electrons \textit{and holes} can be created by optical excitation. By using circularly polarized light it is possible to achieve a high degree of spin polarization of the electrons in the conduction band (the spins of the holes relax rapidly due to strong spin-orbit interactions in the valence band), and the time evolution of this spin polarization can be monitored in real time. Such a system is virtually impurity free, and thus offers a unique opportunity to directly test the impact of many-body interactions on DP spin relaxation of electrons. 

The initial motivation for the present study came from an experiment in which a density-dependent diffusion rate for spin polarized electrons was observed in a photo-excited electron-hole packet in bulk GaAs.\cite{prb_79_115321} Zhao \textit{et al.} found that the spin density in such a packet has a smaller diffusion constant than the carrier density. It was argued that this could be explained by a density dependent spin relaxation rate, where the electron spins in low density regions of the packet relax faster than in the high density regions, leading to the \textit{appearance} of slow diffusion. This behavior is consistent with that of DP spin relaxation controlled by electron-electron scattering in the non-degenerate regime,\cite{jetp_99_6_1279} which is indeed expected to occur in the low density regions of the electron-hole packet. Remarkably, the \textit{increase} of the spin-relaxation rate with decreasing density in the non-degenerate regime is opposite to the behavior in the degenerate regime,\cite{prb_79_125206} where the spin relaxation rate \textit{decreases} with decreasing density. This means that the SRT has a maximum as a function of density at a density intermediate between the degenerate and the non-degenerate regime.

The present paper continues our efforts to develop a better microscopic understanding of the GI mechanism of DP spin relaxation. We consider electron-electron, electron-impurity, and electron-hole interactions in the degenerate regime. In this regime, the presence of a large Fermi surface paves the way to an elegant analytical treatment of electron-electron and electron-hole scattering rates -- a treatment pioneered by Abrikosov and Khalatnikov in their classic paper on the theory of the Fermi liquid. The relevant scattering rates are similar to but not identical with the well-known momentum scattering rates which control, for example, the lifetime of quasiparticles. The difference arises because different collision events are not equally effective at changing the magnitude and direction of the spin precession: the contribution of different collision events must be weighted according to their effectiveness at changing the spin precession. (A similar ``weighting" -- this time concerning the effectiveness of collisions at changing the current -- is responsible for the replacement of the scattering cross section by the ``transport cross section" in microscopic calculations of the electrical conductivity). GI had previously looked at the contribution of electron-electron collisions to DP spin relaxation in the degenerate regime, but limited themselves to qualitative estimates.\cite{jetp_99_6_1279} Jiang, \textit{et al.} applied the powerful kinetic spin Bloch equation approach, including all relevant spin relaxation mechanisms, but their work was primarily numerical.\cite{prb_79_125206} In the present paper, analytic expressions are derived which are applicable to a variety of semiconductors of the zincblende structure; in particular, the calculation of the relevant electron-electron and electron-hole scattering rates is reduced to the evaluation of simple two-dimensional and one-dimensional integrals over the Fermi surface.
We find that electron-hole scattering is the dominant mechanism of momentum randomization in these intrinsic semiconductors. Nevertheless, we emphasize that our mechanism of spin relaxation remains conceptually distinct from the BAP mechanism in which electrons transfer their spin polarization to the holes via interband matrix elements of the Coulomb interaction. In the present mechanism -- that is in the DP mechanism -- electrons transfer only momentum to the holes.

Another novel feature of the present work is that we include exchange and correlation effects in the calculation of the effective electron-electron and electron-hole interactions. In practice, this means that we are going beyond the basic RPA-screened interaction. Exchange and correlation effects are included via local field factors, and we will show that these effects cause an enhancement of the scattering rate -- and a corresponding reduction of spin relaxation rate -- at low density and low temperature (still in the degenerate regime). In the opposite limit of high density, the scattering lifetime becomes very large. A cautionary note is included for the treatment of DP spin relaxation in this regime. Specifically, the common assumption that momentum-changing collisions are frequent on the scale of spin relaxation fails, since the two processes occur on comparable time scales. In this regime the quasiparticles are essentially non-interacting and the DP mechanism is superseded by the spin precession of individual quasiparticles of essentially constant momentum.

This paper is organized as follows: in Section \ref{srt_review}, we review the steps and the assumptions in the derivation of the standard formula for D'yakonov-Perel' spin relaxation; in Section \ref{eff_scat} we derive analytic expressions for the effective (i.e. weighted) scattering rates due to electron collisions with electrons, holes and impurities; Section \ref{scat} describes the effective scattering amplitudes used in our calculations; Section \ref{disc} discusses the effective scattering rates and SRTs for doped and intrinsic GaAs; Section \ref{conc} contains our concluding remarks.\\

\section{D'yakonov-Perel' spin relaxation} \label{srt_review}
Although the derivation of DP spin relaxation is well known and can be found in excellent reviews \cite{opt_orient,fabian_aps} we briefly reproduce it here, partly to adapt the notations, and partly to spell out the underlying physical assumptions. The system under investigation is a III-V semiconductor (i.e. GaAs, InAs, InSb) of the zibcblende structure, which exhibits spin-split bands. The spin splitting is caused by a Dresselhaus effective magnetic field\cite{pr_100_580} which arises from spin-orbit interaction and couples to electrons via the hamiltonian term 
\begin{equation} \label{H1k}
	H_{1\kv} = \frac{\hbar}{2} \Ov_{\kv} \cdot \sigv \, ,
\end{equation}
where $\Ov_{\kv}$ defines the precession axis and precession frequency of electron spins. $\Ov_{\kv}$ is a cubic harmonic function of the Bloch wave vector $\kv$; its component along the $\zh$-axis is given by
\begin{equation} \label{Omega}
	\Omega_{\kv,z} = \frac{\alpha_c \hbar^2}{\sqrt{2m_c^3E_g}} k_z(k_x^2-k_y^2) \, ,
\end{equation}
where $\alpha_c$ is the Dresselhaus spin-orbit coupling constant (e.g. $\alpha_c=0.07$ for GaAs), $E_g$ is the band gap energy, and $m_c$ is the conduction band effective mass of an electron. The other components of $\Ov_{\kv}$ are obtained by cyclic permutations of $x$, $y$, and $z$. 

The spin Boltzmann kinetic equation describes the evolution of electrons in time:
\begin{equation} \label{sBe}
	\frac{\partial \rho_{\kv}}{\partial t} - \frac{1}{i\hbar}[H_{1\kv},\rho_{\kv}] + \frac{\partial \rho_{\kv}}{\partial \hbar\kv} \cdot \mathbf{F_k} + \frac{\partial \rho_{\kv}}{\partial \mathbf{r}} \cdot \mathbf{v}_{\kv} = I_{\kv}(t)\, ,
\end{equation}
where $\rho_{\kv}$ is the $2\times 2$ spin density matrix and $I_{\kv}(t)$ is the collision integral (also a $2\times 2$ matrix). Assuming a homogeneous distribution of electrons ($\partial \rho_{\kv}/\partial \mathbf{r} = 0$) and absence of external fields ($\mathbf{F_k}=0$), Eq.~\eqref{sBe} reduces to
\begin{equation} \label{sBe_reduced}
	\frac{\partial \rho_{\kv}}{\partial t} - \frac{1}{2i}[\Ov_{\kv}\cdot\sigv,\rho_{\kv}] = I_{\kv}(t)\, .
\end{equation}

We prepare a spin polarized distribution which evolves according to Eq.~\eqref{sBe_reduced} and eventually relaxes to a completely unpolarized state. To describe the process, we search for a solution which contains a quasi-equilibrium component, $\rho_{0k}$, describing a state of uniform spin polarization along a direction denoted by $\sh$, and a non-equilibrium component, $\rho_{1\kv}$:
\begin{equation} \label{rho_decomp}
	\rho_{\kv} = \rho_{0k} + \rho_{1\kv} \, .
\end{equation}
Using this definition in Eq.~\eqref{sBe_reduced}, we can relate terms of varying order of spin orbit interaction ($\alpha_c$):
\begin{multline} \label{sBe_expanded}
	\frac{\partial \rho_{0k}}{\partial t} + \frac{\partial \rho_{1\kv}}{\partial t} - \frac{1}{2i}[\Ov_{\kv}\cdot\sigv,\rho_{0k}] \\
	- \frac{1}{2i}[\Ov_{\kv}\cdot\sigv,\rho_{1\kv}] = I_{\kv}(t)\, .
\end{multline}
The following presumptions are made about the various terms appearing in Eq.~(\ref{sBe_expanded}):
\begin{enumerate}
	\item Order of spin-orbit interaction ($\alpha_c$): \\
		  $\rho_{0k}$: $0^{\text{th}}$ order; $\rho_{1\kv}$: $1^{\text{st}}$ order; \\
		  $\dot{\rho}_{0k}$: $2^{\text{nd}}$ order; $\dot{\rho}_{1\kv}$: $3^{\text{rd}}$ order.
	\item $I_{\kv}(t)$ may be cast in the form of relaxation time approximation.
\end{enumerate}
The first of these presumptions will be shown to be consistent with the form of the kinetic equation momentarily. The legitimacy of the second presumption will be discussed in the next section when we analyze the collision integral in detail. For now, we simply write
\begin{equation} \label{RTA}
	I_{\kv}(t) = -\frac{\rho_{1\kv}}{\tau^*_{\kv}}\, ,
\end{equation}
where $\tau^*_{\kv}$ is an effective scattering time on the order of the plane wave lifetime. The precise form of $\tau^*_{\kv}$ depends on the scattering mechanism and will be explicitly constructed in the next sections for various cases of interest. Notice that $\rho_{0k}$, being a quasi-equilibrium distribution, does not contribute to the collision integral.

Equating terms of the same order in $\alpha_c$ in Eq.~\eqref{sBe_expanded}, the following relations emerge:
\begin{align}
	\label{r1k_relation} I_{\kv}(t) &= -\frac{\rho_{1\kv}}{\tau^*_{\kv}} = -\frac{1}{2i}[\Ov_{\kv}\cdot\sigv,\rho_{0k}] \\
	\label{r0k_relation} \frac{\partial \rho_{0k}}{\partial t} &= \frac{1}{2i}\langle{[\Ov_{\kv}\cdot\sigv,\rho_{1\kv}]}\rangle_k \\
	\frac{\partial \rho_{1\kv}}{\partial t} &= 0 \, .
\end{align}
where $\langle\dots\rangle_k$ stands for the angular average over all directions of $\kv$ at fixed magnitude $k=|\kv|$. The angular averaging is necessary in Eq.~\eqref{r0k_relation} to properly relate the quantities. Finally, $\partial \rho_{1\kv}/\partial t$ has no counterpart in the kinetic equation at order $\alpha_c^2$, hence it must vanish at this order as anticipated. 

Let us now introduce the form of the quasi-equilibrium distribution, $\rho_{0k}$, from which we may use the relations above to find expressions for $\rho_{1\kv}$ and the spin polarization. $\rho_{0k}$ is written in terms of equilibrium Fermi-Dirac distributions for spin-up electrons ($f^{\ua}$) and spin-down electrons ($f^{\da}$) in the presence of a fictitious $k$-dependent magnetic field which we introduce to enforce the desired quasi-equilibrium spin polarization:
\begin{equation} \label{rho_avg}
	\rho_{0k} = f^n_k \one + f^s_k \sh \cdot \sigv \, ,
\end{equation} 
where $\one$ is the $2\times 2$ identity matrix,
\begin{equation}
	f^n_k = \frac{f^{\ua}_k + f^{\da}_k}{2}
\end{equation}
is the spin-averaged occupation number of the $\kv$ state and
\begin{equation}
	f^s_k = \frac{f^{\ua}_k - f^{\da}_k}{2}
\end{equation}
is the quasi-equilibrium spin polarization occupation number of the $\kv$ state.
The distributions for spin-up and spin-down electrons are defined by
\begin{align}
	f^{\ua/\da}_k &= \frac{1}{1+e^{\beta(\ve_k-\mu\mp\ve_s)}} \, ,
\end{align}
where $\beta = 1/(k_BT)$, $\mu$ is the chemical potential, and $\ve_s$ is the Zeeman interaction energy.

The non-equilibrium part of the distribution function arises from the competition between spin precession, which tilts the spins away from the $\sh$-axis, and collisions, which attempt to restore local equilibrium. Inserting $\rho_{0k}$ [Eq.~\eqref{rho_avg}] into the relation for $\rho_{1\kv}$ [Eq.~\eqref{r1k_relation}], we find
\begin{equation}
	\label{rho1k} \rho_{1\kv} = \tau^*_{\kv} \l(\Ov_{\kv}\times f^s_k \sh\r) \cdot \sigv \, .
\end{equation}
At this point, the presumption made earlier in this section concerning the orders of $\alpha_c$ in $\rho_{0k}$, $\rho_{1\kv}$, $\dot{\rho}_{0k}$, and $\dot{\rho}_{1\kv}$ may be verified easily.

We identify the quasi-equilibrium spin polarization at wave vector $k$ as
\begin{equation} \label{Def_sv}
	\sv_{0k} = \frac{\hbar}{2}\Tr[\rho_{0k} \sigv] = \hbar f^s_k \sh \,
\end{equation}
and notice that it does not depend on the direction of $\kv$. Then, the time rate of change of $\sv_{0k}$ is found by tracing Eq.~\eqref{r0k_relation} with $\sigv$:
\begin{equation} \label{sdot_in_crosses}
	\frac{\partial \sv_{0k}}{\partial t} = \langle{\tau^*_{\kv} \Ov_{\kv}\times \l(\Ov_{\kv}\times\sv_{0k}\r)}\rangle_k \, .
\end{equation}
This equation can be cast into the standard relaxation-type form,
\begin{equation}
	\label{bloch_srt} \frac{\partial \sv_{0k}}{\partial t} = - \frac{\sv_{0k}}{\tau_k^{(s)}}\, ,
\end{equation}
by defining the rate of spin relaxation at $k$ as follows:
\begin{equation} \label{tau_s_simple}
	\frac{1}{\tau_k^{(s)}} = \langle{\tau_{\kv}^* \l( \Omega_{\kv}^2 - (\Ov_{\kv}\cdot\sh)^2 \r)}\rangle_k \, .
\end{equation}

The cubic symmetry of III-V semiconductors implies that (i) the relaxation rate of Eq.~(\ref{tau_s_simple}) is independent of the direction of $\sh$, and (ii) the effective collision time $\tau_{\kv}^*$ depends on $\kv$ only through its modulus $k$ (in the next sections we verify this explicitly). Therefore, for cubic systems, the expression for the spin relaxation rate (\ref{tau_s_simple}) simplifies further
\begin{equation}
\label{tau_s_simple2}
	\frac{1}{\tau^{(s)}_k} = \frac{2}{3}\tau^*_k \langle{\Omega_{\kv}^2}\rangle_k \, .
\end{equation}

The experimentally measurable quantity of primary physical interest is the expectation value of the spin density
\begin{equation}
	\Sv = \sum_{\kv} \sv_{0k} \, .
\end{equation}
The equation of motion for $\Sv(t)$ is obtained by summing Eq.~\eqref{sdot_in_crosses} over all $\kv$ and using this result in the relaxation time approximation defining $\tau^{(s)}$:
\begin{equation}
	\frac{\partial \Sv}{\partial t} = - \frac{\Sv}{\tau^{(s)}}\, ,
\end{equation}
where the physical spin relaxation time is calculated from
\begin{equation} \label{tau_s_two_sums}
	\frac{1}{\tau^{(s)}} = \frac{2}{3}\frac{\sum_{\kv}\tau^*_k f^s_k\langle{\Omega_{\kv}^2}\rangle_k }{\sum_{\kv} f^s_k} \, .
\end{equation}

Obviously, the spin relaxation time as an internal material parameter, independent of a particular non-equilibrium state, is only meaningful in the regime of small spin polarization. In this limit, we can linearize $f^s_k$ in the polarization energy, $\ve_s$:
\begin{equation}
\label{f-linear}
	f^s_k = -\ve_s\frac{\partial f^0_k}{\partial\ve_k} = \beta\ve_s f^0_k(1-f^0_k) \, ,
\end{equation}
where $f^0_k$ is the unpolarized Fermi-Dirac function. Substituting Eq.~(\ref{f-linear}) into Eq.~(\ref{tau_s_two_sums}) we indeed find that the polarization energy, $\ve_s$, cancels and we are left with the following expression which contains only internal characteristics of the system:
\begin{equation}
 \label{tau-fin}
\frac{1}{\tau^{(s)}} = \frac{2}{3}\frac{\sum_{\kv}\tau^*_k f^0_k(1-f^0_k)\langle{\Omega_{\kv}^2}\rangle_k}{\sum_{\kv} f^0_k(1-f^0_k)} \, .
\end{equation}
The spin relaxation time $\tau^{(s)}$ defined by this equation is the experimentally accessible quantity which determines a universal long time tail, $\Sv(t)\sim e^{-t/\tau^{(s)}}$, in the spin relaxation dynamics.

In the degenerate limit, $k_BT\ll\ve_F$, the factor $f^0_k(1-f^0_k)$ entering the intergals in Eq.~(\ref{tau-fin}) is strongly peaked at the Fermi level. Since the spin orbit field $\Omega_{\kv}$ in Eq.~(\ref{Omega}) is a slowly varying function of $k$, it is legitimate to set $k=k_F$ and take it out of the integral. As a result, the spin relaxation rate takes the following simple form
\begin{equation}
\label{tau_fin-degen}
	\frac{1}{\tau^{(s)}} = \frac{2}{3}\tau^*_{\text{avg}} \langle{\Omega_{\kv}^2}\rangle_{k_F} \, ,
\end{equation}
where
\begin{equation}
 \label{tau-avg1}
\tau^*_{\text{avg}} = \frac{\sum_{\kv}\tau^*_k f^0_k(1-f^0_k)}{\sum_{\kv} f^0_k(1-f^0_k)} \, .
\end{equation}
Hence, in the degenerate limit, the spin relaxation time $\tau^{(s)}$ is completely determined by two parameters: the mean square of the spin-orbit field $\langle{\Omega_{\kv}^2}\rangle_{k_F}$ at the Fermi energy and a properly averaged effective scattering time $\tau^*_{\text{avg}}$. While calculation of $\langle{\Omega_{\kv}^2}\rangle_{k_F}$ is straightforward, to find $\tau^*_{\text{avg}}$ we normally need to solve a complicated integral equation. One of the goals of this paper is to show that $\tau^*_{\text{avg}}$, and thus $\tau^{(s)}$, can be calculated rigorously using methods developed in the theory of Fermi liquids.\cite{SykesBrooker}

There are two important points to be made about the above derivation. First, the whole treatment is reliant on the spin-orbit interaction being weak, leading to a clear separation of time scales between the microscopic momentum changing-collisions (fast) and the macroscopic spin relaxation (slow). In other words, we must have $\Omega_{\kv}\tau^*_{\kv} \ll 1$; spins relax due to a combination of $\kv$-randomizing collisions and relatively small precession rotations. From here, we can see that when the scattering lifetime is of the same order of magnitude as the spin-relaxation time, our treatment is no longer justified. If spins relax just as quickly as collisions occur, it makes no sense to introduce a quasi-equilibrium distribution function. In the extreme limit of infrequent collisions, the momentum of the electron becomes a constant of the motion and the spin simply precesses in the Dresselhaus field at a given $\kv$.

The second point is that Eq.~\eqref{tau_s_simple} is often estimated by replacing the effective scattering time $\tau^*_{\kv}$ with the plane wave lifetime $\tau_{\kv}$. This, however, is not correct, for the reasons described in the introduction. $\tau^*_{\kv}$ contains an additional momentum axis re-orientation mechanism, due to the Dresselhaus field. In the next section we will obtain the correct expression for $\tau^*_{\kv}$, and in Section \ref{disc} we will explore the validity of the approximate replacement of $\tau^*_{\kv}$ by $\tau_{\kv}$.

\section{The effective scattering time} \label{eff_scat}
In order to calculate the effective scattering time $\tau_{\kv}^*$ we need to construct the collision integral. Naturally, the construction depends on the nature of the collision process. Let us begin with electron-electron scattering processes. Electron-hole scattering processes will then be easily handled. These two scattering mechanisms are all that is needed to treat DP spin relaxation in intrinsic, photo-excited semiconductors. For doped semiconductors, electron-impurity scattering needs to be included.

\subsection{Electron-electron collisions}

The collision integral for an interacting Fermi liquid has been derived by several authors.\cite{jpf_49_10_1691,prb_69_144429,jetp_99_6_1279} The most direct derivation starts from the Kadanoff-Baym quantum kinetic equations.\cite{haug_qkt} The collision integral derived in this manner has the form
\begin{equation}
	I_{\kv}(t) = I_{\kv}^{\text{in}}(t)-I_{\kv}^{\text{out}}(t) \, ,
\end{equation}
where
\begin{subequations} \label{Collision_Integral}
\begin{multline}
	I_{\kv}^{\text{in}}(t) = \int_{-\infty}^t dt_1 \l\{\Sigma^<_{\kv}(t,t_1) G^>_{\kv}(t_1,t) \r. \\
	\l. + G^>_{\kv}(t,t_1)\Sigma^<_{\kv}(t_1,t)\r\} \, , 
\end{multline}
\begin{multline}
	I_{\kv}^{\text{out}}(t) = \int_{-\infty}^t dt_1 \l\{\Sigma^>_{\kv}(t,t_1) G^<_{\kv}(t_1,t) \r. \\
	\l. + G^<_{\kv}(t,t_1)\Sigma^>_{\kv}(t_1,t)\r\} \, .
\end{multline}
\end{subequations}
\begin{figure*}[bht]
\begin{center}
	\includegraphics[width=15cm]{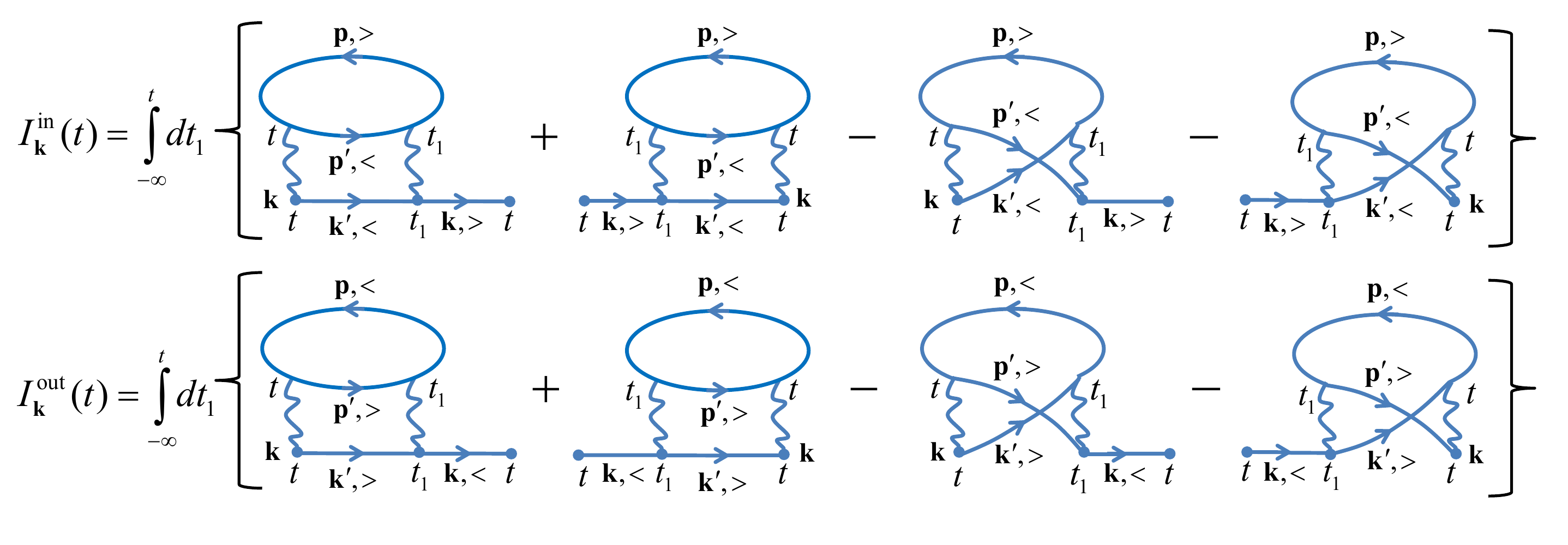}
\end{center}
\caption[Collision Integral]{(Color online) Contributions to the collision integral under the Born approximation.}
\label{fig:coll_int_diag}
\end{figure*}
Here, the lesser and greater Green's functions and the corresponding self-energies are all $2\times 2$ matrices in spin space. For the self-energy function one adopts the Born approximation, in other words, second order in an effective electron-electron interaction. The scattering amplitude is denoted by $W_{\kv\pv\kv'\pv'}$, where $\kv,\pv$ are the momenta of the incoming particles and $\kv',\pv'$ those of the outgoing ones (Notice that one must have $\kv+\pv=\kv'+\pv'$ by momentum conservation). In the simplest approximation $W$ is simply the Fourier transform of the Coulomb interaction:
\begin{equation}
	W_{\kv\pv\kv'\pv'} = \frac{4\pi e^2}{\epsilon|\kv-\kv'|^2} \delta_{\kv+\pv,\kv'+\pv'}\, ,
\end{equation}
where $\epsilon$ is the background dielectric constant of the semiconductor and $e$ is the electron charge. Here, we will include RPA screening and post-RPA screening via local field factors. In a more complete treatment, the effective interaction would also depend on the relative orientation of the spins; we will not consider such effects here, but simply assume that $W$ is averaged over the relative spin orientations.

In Fig.~\ref{fig:coll_int_diag} we show the diagrams that contribute to the collision integral in this approximation. As a further approximation we adopt the generalized Kadanoff-Baym (GKB) ansatz\cite{PhysRevB.34.6933,haug_qkt} relating the Green's functions to the distribution function:
\begin{equation}
G^<_{\kv}(t,t')=\rho_{\kv}(t)G^a_{\kv}(t,t')-G^r_{\kv}(t,t')\rho_{\kv}(t')\, ,
\end{equation}
where
\begin{subequations}
\begin{align}
	G^r_{\kv}(t,t') &= -i\theta(t-t')e^{-i \ve_k (t-t')}\, ,\\
	G^a_{\kv}(t,t') &= i\theta(t'-t)e^{i \ve_k (t-t')}\, ,
\end{align}
\end{subequations}
and $\ve_k = \hbar^2k^2/2m^*$ is the single-electron energy in a parabolic band of effective mass $m^*$. Notice that we are neglecting interaction contributions to the single particle energy, coming, for example, from exchange. It is assumed that these contributions are properly included in the effective mass. As a final approximation we ignore the difference between $\rho_{\kv}(t_1)$ and $\rho_{\kv}(t)$, i.e. we assume that the distribution function is slowly varying on the time scale probed by the integrals in Eq.~\eqref{Collision_Integral}. This is commonly referred to as the Markovian approximation, and is completely justified in the solution of a steady-state problem. Then, the remaining integral over time contains only a principle value term and a $\delta$ of conservation of energy. Retaining only the dissipative contribution (from the conservation of energy term), the final expression is
\begin{widetext}
\begin{multline} \label{coll_int_ee}
	I_{\kv}(t)^{\text{e-e}} = -\frac{\pi}{\hbar} \sum_{\kv' \pv \pv'} \delta\l(\ve_k+\ve_p-\ve_{k'}-\ve_{p'}\r) \delta_{\kv+\pv,\kv'+\pv'} \\
	\l[ \l|W_{\kv \pv \kv' \pv'}\r|^2 \l( \l\{\rho_{\kv},\one-\rho_{\kv'}\r\} \Tr \l[\rho_{\pv}\l(\one-\rho_{\pv'}\r)\r] - \l\{\one-\rho_{\kv},\rho_{\kv'}\r\} \Tr \l[ \l(\one-\rho_{\pv}\r)\rho_{\pv'} \r] \r) \r. \\
	\l. -\l|W_{\kv \pv \kv' \pv'}W_{\kv \pv \pv' \kv'}\r| \l( \l\{\rho_{\kv},\l(\one-\rho_{\kv'}\r)\rho_{\pv}\l(\one-\rho_{\pv'}\r) \r\} - \l\{\one-\rho_{\kv},\rho_{\kv'}\l(\one-\rho_{\pv}\r)\rho_{\pv'}\r\} \r) \r] \, .
\end{multline}
\end{widetext}

We now show that the relaxation-time approximation in Eq.~\eqref{RTA} is consistent with this collision integral linearized in $\rho_{1\kv}$. With $\delta\l(\ve_k+\ve_p-\ve_{k'}-\ve_{p'}\r)$, the quasi-equilibrium distributions make no contribution:
\begin{multline} \label{fd_id1}
	\l\{\rho_{0k},\l(\one-\rho_{0k'}\r)\r\}\Tr\l[\rho_{0p}\l(\one-\rho_{0p'}\r)\r] = \\
	\l\{\l(\one-\rho_{0k}\r),\rho_{0k'}\r\}\Tr\l[\l(\one-\rho_{0p}\r)\rho_{0p'}\r]
\end{multline}
and
\begin{multline} \label{fd_id2}
	\l\{\rho_{0k},\l(\one-\rho_{0k'}\r)\rho_{0p}\l(\one-\rho_{0p'}\r)\r\} = \\
	\l\{\l(\one-\rho_{0k}\r),\rho_{0k'}\l(\one-\rho_{0p}\r)\rho_{0p'}\r\} \, .
\end{multline}
Then, each term in the collision integral contains one $\rho_1$:
\begin{equation} \label{linear_direct}
\begin{split}
	&\l\{\rho_{\kv},\one-\rho_{\kv'}\r\} \Tr \l[\rho_{\pv}\l(\one-\rho_{\pv'}\r)\r] \\
	-&\l\{\one-\rho_{\kv},\rho_{\kv'}\r\} \Tr \l[\l(\one-\rho_{\pv}\r) \rho_{\pv'}\r] = \\
	&\quad\quad \l\{\rho_{1\kv},\one-\rho_{0k'}\r\}\Tr\l[\rho_{0p}\l(\one-\rho_{0p'}\r)\r] \\
	&\quad\quad + \l\{\rho_{1\kv},\rho_{0k'}\r\}\Tr\l[\l(\one-\rho_{0p}\r)\rho_{0p'}\r] \\
	&\quad\quad - \l\{\rho_{0k},\rho_{1\kv'}\r\}\Tr\l[\rho_{0p}\l(\one-\rho_{0p'}\r)\r] \\
	&\quad\quad - \l\{\one-\rho_{0k},\rho_{1\kv'}\r\}\Tr\l[\l(\one-\rho_{0p}\r)\rho_{0p'}\r] \\
	&\quad\quad + \l\{\rho_{0k},\one-\rho_{0k'}\r\}\Tr\l[\rho_{1\pv}\l(\one-\rho_{0p'}\r)\r] \\
	&\quad\quad + \l\{\one-\rho_{0k},\rho_{0k'}\r\}\Tr\l[\rho_{1\pv}\rho_{0p'}\r] \\
	&\quad\quad - \l\{\rho_{0k},\one-\rho_{0k'}\r\}\Tr\l[\rho_{0p}\rho_{1\pv'}\r] \\
	&\quad\quad - \l\{\one-\rho_{0k},\rho_{0k'}\r\}\Tr\l[\l(\one-\rho_{0p}\r)\rho_{1\pv'}\r]
\end{split}
\end{equation}
and
\begin{equation} \label{linear_exchange}
\begin{split}
	&\l\{\rho_{\kv},\l(\one-\rho_{\kv'}\r) \rho_{\pv}\l(\one-\rho_{\pv'}\r)\r\} \\
	-&\l\{\one-\rho_{\kv},\rho_{\kv'}\l(\one-\rho_{\pv}\r)\rho_{\pv'}\r\} = \\
	&\quad\quad\l\{\rho_{1\kv},\l(\one-\rho_{0k'}\r)\rho_{0p}\l(\one-\rho_{0p'}\r)\r\} \\
	&\quad\quad+ \l\{\rho_{1\kv},\rho_{0k'} \l(\one-\rho_{0p}\r)\rho_{0p'}\r\} \\
	&\quad\quad- \l\{\rho_{0k},\rho_{1\kv'} \rho_{0p}\l(\one-\rho_{0p'}\r)\r\} \\
	&\quad\quad- \l\{\one-\rho_{0k},\rho_{1\kv'} \l(\one-\rho_{0p}\r)\rho_{0p'}\r\} \\
	&\quad\quad+ \l\{\rho_{0k},\l(\one-\rho_{0k'}\r) \rho_{1\pv}\l(\one-\rho_{0p'}\r)\r\} \\
	&\quad\quad+ \l\{\one-\rho_{0k},\rho_{0k'} \rho_{1\pv}\rho_{0p'}\r\} \\
	&\quad\quad- \l\{\rho_{0k},\l(\one-\rho_{0k'}\r) \rho_{0p}\rho_{1\pv'}\r\} \\
	&\quad\quad- \l\{\one-\rho_{0k},\rho_{0k'} \l(\one-\rho_{0p}\r)\rho_{1\pv'}\r\} \, .
\end{split}
\end{equation}
These can be simplified a bit further, but at a loss of readability.

We assumed in Eq.~\eqref{r1k_relation} that the collision integral is proportional to $\rho_{1\kv}$, which is in turn proportional to $\l[\Ov_{\kv}\cdot\sigv , \rho_{0k}\r]$. Furthermore, each component of $\Ov_{\kv}$ gets all its angular dependencies from $l=3$ spherical harmonics [e.g. $\Omega_{\kv,z}$ is expressed this way in Eq.~\eqref{omega_z}]. Obviously $\rho_{0k}$ contains no such terms. Therefore $\rho_{1\kv}$ must contain only $l=3$ spherical harmonics. Viewing the collision integral as an integral operator,
\begin{equation}
	\El \rho_{1\kv} = \sum_{\kv'} \El(\kv,\kv') \rho_{1\kv'} \, ,
\end{equation}
it is apparent that $\El$, being rotationally invariant, has no means of changing the harmonic content of $\rho_{1\kv'}$. Then, we reach the conclusion that the collision integral also contains only $l=3$ harmonics. In other words, we see that the relaxation time approximation introduced in Eq.~\eqref{RTA} is consistent with the form of the linearized collision integral.

Clearly, $\rho_{1\kv}$ is traceless; so then is the collision integral. We trace the linearized collision integral with $\sigv$ to find a calculable relationship between elements in $\rho_{1\kv}$ and $I_{\kv}$. In the limit of small spin polarization $\rho_{\kv}$ is expanded to first order in the polarization energy with the assumption that $\ve_s \ll \ve_k$. $\rho_{0k}$ and $\rho_{1\kv}$ then take the forms
\begin{align}
	\rho_{0k} &= f^0_k \one + \beta \ve_s f^0_k(1-f^0_k) \hat{\sv} \cdot \sigv \, , \\
	\rho_{1\kv} &= \tau^*_{\kv} \beta \ve_s f^0_k(1-f^0_k) \l(\Ov_{\kv} \times \hat{\sv}\r)\cdot\sigv \, ,
\end{align}
where $f^0_k $ is the unpolarized Fermi-Dirac distribution. Equating terms of first order in $\ve_s$ we find
\begin{widetext}
\begin{multline} \label{coll_int_reduced}
	f^0_k (1-f^0_k)\Ov_{\kv}\times\sh = \frac{4\pi}{\hbar} \frac{1}{\l(2\pi\r)^6} \sum_{\pv'} \int d\kv'\, d\pv\, f^0_k(1-f^0_{k'})f^0_p(1-f^0_{p'}) \delta\l(\ve_k+\ve_p-\ve_{k'}-\ve_{p'}\r) \delta_{\kv+\pv,\kv'+\pv'} \\
	\l\{ \l|W_{\kv \pv \kv' \pv'}\r|^2 \l[\tau^*_{\kv} \Ov_{\kv} - \tau^*_{\kv'} \Ov_{\kv'} \r] - \frac{1}{2}\l|W_{\kv \pv \kv' \pv'}W_{\kv \pv \pv' \kv'}\r| \l[\tau^*_{\kv} \Ov_{\kv} - \tau^*_{\kv'} \Ov_{\kv'} + \tau^*_{\pv} \Ov_{\pv} - \tau^*_{\pv'} \Ov_{\pv'}\r] \r\} \times\sh \, .
\end{multline}
\end{widetext}

Cubic symmetry in III-V semiconductors permits us to consider only one component of this vector relation, $\Ov_{\kv}\times\sh \rightarrow \Omega_{\kv,z}$. The later quantity can be conveniently written in terms of spherical harmonics, $Y_l^m(\theta,\phi)$:
\begin{equation} \label{omega_z}
	\Omega_{\kv,z} = \frac{\alpha_c \hbar^2 k^3}{\sqrt{2m_c^3E_g}} \sqrt{\frac{8\pi}{105}} \l[Y_3^2(\vt_k,\vp_k)+Y_3^{-2}(\vt_k,\vp_k)\r] \, .
\end{equation}
This fact, together with isotropy of the scattering, imply that the solution $\tau^*_{\kv}$ of Eq.~(\ref{coll_int_reduced}) does not depend on the direction of $k$. Indeed, assuming $\tau^*_{\kv}=\tau^*_{k}$, we find that the presence of $\Omega_{\kv,z}$ [given by Eq.~(\ref{omega_z})] in the integrals in Eq.~(\ref{coll_int_reduced}) already guarantees that the whole right hand side has the proper angular dependences, consistent with that of the left hand side. As further evidence of this fact, it is easily demonstrated that integration of $\Omega_{\qv,z}$ (where $\qv$ can be any of $\kv'$, $\pv$, or $\pv'$) over $d\qh$ results in a term proportional to $\Omega_{\kv,z}$. This is shown in Appendix~\ref{r1x_to_omega_k}.

As usual, further simplifications come in the degenerate limit because the factor $f^0_k(1-f^0_{k'})f^0_p(1-f^0_{p'})$ confines the momentum integrals to a narrow shell around the Fermi energy, where the density of states can be well approximated by a constant. In this case, $\tau^*_{k}=\tau^*(\xi)$ becomes a function of the dimensionless energy variable, $\xi=(\ve_k-\mu)/(k_BT)$, which satisfies a one-dimensional integral equation of the following form:
\begin{equation}
\label{1D-equation}
	B = \int_{-\infty}^{\infty} dx\, K(x,\xi) \l[ \tau^*(\xi) - \lambda\tau^*(x) \r] \, .
\end{equation}
The first term of the integral corresponds to the combination of each of the first terms in square brackets of Eq.~\eqref{coll_int_reduced}; the second term in the integral picks up all of the remaining terms of Eq.~\eqref{coll_int_reduced}. A detailed derivation of this equation, starting with Eq.~(\ref{coll_int_reduced}), is given in Appendix~\ref{int_eqn_reduction}. The kernel $K(x,\xi)$ and the parameters $B$ and $\lambda$ entering Eq.~(\ref{1D-equation}) are defined as follows:
\begin{align}
	K(x,\xi) =& \frac{f^0(-x)}{f^0(-\xi)} \l[\frac{\xi-x}{1-e^{x-\xi}}\r] \label{kernel} \, ,\\
	B =& \frac{\hbar^7(2\pi)^4}{m_c^3 \l(k_BT\r)^2} \l(A_1-\frac{A_2}{2}\r)^{-1} \, , \label{BFull} \\
	\lambda =& \frac{A_1\lambda_1 - A_2\lambda_2/2}{A_1-A_2/2} \, , \label{lambdaFull} \\
	A_1 =& \int d\Omega\, \frac{\l|W_{\kv\kv'}\r|^2}{\cos(\theta/2)} \, , \label{A1} \\
	A_2 =& \int d\Omega\, \frac{\l|W_{\kv\kv'}W_{\kv\pv'}\r|}{\cos(\theta/2)} \label{A2} \, ,\\
	\lambda_1 =& \frac{1}{A_1} \int d\Omega\, \frac{\l|W_{\kv\kv'}\r|^2}{\cos(\theta/2)} P_3\l( \cos\theta_1 \r) \, , \label{lambda1} \\
	\begin{split}
		\lambda_2 =& \frac{1}{A_2} \int d\Omega\, \frac{\l|W_{\kv\kv'}W_{\kv\pv'}\r|}{\cos(\theta/2)} \l[P_3\l( \cos\theta_1 \r) \r. \\
		& \l. - P_3\l(\cos\theta \r) + P_3\l( \cos\theta_2 \r)\r] \, . \label{lambda2}
	\end{split}
\end{align}
$P_3(x)$ is the 3rd order Legendre polynomial. The solid angle $d\Omega=\sin\theta\, d\theta d\phi$ should not be confused with the Dresselhaus Larmor frequency. All of these angles ($\theta,\phi,\theta_1,\theta_2$) are described in Appendix \ref{int_eqn_reduction}, but briefly: $\theta$ is the angle between $\kv$ and $\pv$; $\phi$ is the polar angle about the $\kv+\pv$ axis, between $\kv'$ ($\pv'$) and $\kv$ ($\pv$); $\cos\theta_1 = \kh\cdot\kh' = (1/2)(1 + \cos\theta + \cos\phi - \cos\theta\cos\phi)$; $\cos\theta_2 = \kh\cdot\ph' = (1/2)(1 + \cos\theta - \cos\phi + \cos\theta\cos\phi)$.

Integral equations of the type in Eq.~(\ref{1D-equation}) with a kernel of Eq.~(\ref{kernel}) are common in the theory of transport coefficients of Fermi liquids .\cite{abrikosov_tfl,baym_pethick_lflt} The general method of solution has been proposed by Sykes and Brooker.\cite{SykesBrooker} This amounts to a clever Fourier transform utilizing the convolution theorem and ultimately changing the integral equation to a recognizable inhomogeneous differential equation. The solution to our Eq.~(\ref{1D-equation}) may then literally be read out from their paper:
\begin{equation} \label{tau_star_k_final}
	\tau^*(\xi) = \frac{\cosh(\xi/2)}{2\pi}\int_{-\infty}^{\infty} d\omega\, e^{-i\omega \xi} \frac{B}{\pi} \sum_{l=0}^{\infty} \frac{(4l+3)\Phi_{2l}(\omega)}{\Lambda_{2l}\l(\Lambda_{2l} - \lambda\r)} \, ,
\end{equation}
where
\begin{align}
	\Phi_l(\omega) &= p_{l+1}^1 \l(\tanh \pi \omega\r)\, , \\
	\Lambda_l &= \frac{1}{2}(l+1)(l+2)\, ,
\end{align}
and $p_l^m(x)$ are the associated Legendre polynomials.

In Section \ref{srt_review} we showed that the physical spin relaxation time is proportional to an average effective scattering time $\tau^*_{\text{avg}}$ given by Eq.~(\ref{tau-avg1}). Expressing the summation over $\kv$ in Eq.~(\ref{tau-avg1}) in terms of a $\xi$-integral and using the solution of Eq.~(\ref{tau_star_k_final}), we arrive at our final result:
\begin{align} \label{scat_avg_ee}
\begin{split}
	\tau^*_{\text{avg}} &= \int_{-\infty}^{\infty} d\xi \l(-\frac{d f^0_k}{d\xi}\r) \tau^*(\xi) \\
	&= \frac{B}{2\pi^2} \sum_{l=0}^{\infty} \frac{4l+3}{\Lambda_{2l}\l[\Lambda_{2l}-\lambda\r]} \, .
\end{split}
\end{align}

As it turns out, for all calculations performed in this paper, the first two terms of the sum account for greater than 99\% of total scattering time. The first term alone is about 95\% of the sum at the lowest density examined ($n \sim 10^{16}$cm$^{-3}$) and the accuracy increases with increasing density.

For reference, we also mention the plane wave scattering time. This can be calculated readily from Eq.~\eqref{final_scat_ak} by setting $P_3(x) \rightarrow 0$ and $\tau^*_k \rightarrow \tau_k$. The $k$-dependent scattering time is
\begin{equation}
	\tau_k = 2B \l[\l(\frac{\ve_k-\mu}{k_BT}\r)^2+\pi^2\r]^{-1} \, 
\end{equation}
and the averaged scattering time is
\begin{equation}
	\tau_{\text{avg}} = \frac{3B}{2\pi^2} \, .
\end{equation}
Notice the close relationship between the plane wave scattering time and the $l=0$ term of the effective scattering time:
\begin{equation}
	\tau^*_{\text{avg},l=0} = \frac{3B}{2\pi^2}\l(\frac{1}{1-\lambda}\r) \, .
\end{equation}
For $0<\lambda<1$, the effective scattering time is enhanced (or the rate is decreased) compared to the plane wave scattering time. This range of $\lambda$ is consistent with the calculations we make including the Dresselhaus field in Section \ref{disc}.

\subsection{Electron-hole collisions}
The collision integral for electron-hole collisions is simply the direct-process-only portion of Eq.~\eqref{coll_int_ee}:
\begin{multline} \label{coll_int_eh}
	I_{\kv}(t)^{\text{e-h}} = \\
	-\frac{\pi}{\hbar} \sum_{\kv' \pv \pv'} \delta\l(\ve_k+\ve_p-\ve_{k'}-\ve_{p'}\r) \delta_{\kv+\pv,\kv'+\pv'} \l|W_{\kv\kv'}\r|^2 \\
	 \l( \l\{\rho_{\kv},\one-\rho_{\kv'}\r\} \Tr \l[\rho_{\pv}^{(h)}\l(\one-\rho_{\pv'}^{(h)}\r)\r] \r. \\
	- \l. \l\{\one-\rho_{\kv},\rho_{\kv'}\r\} \Tr \l[ \l(\one-\rho_{\pv}^{(h)}\r) \rho_{\pv'}^{(h)} \r] \r)\, ,
\end{multline}
where $\rho^{(h)}$ is the hole density matrix. We consider equal densities of electrons and holes so that the Fermi momentum wavevectors are equal for the two species. This condition is likened to an intrinsic semiconductor under optical excitation. Additionally, the valence band energy for holes is considered parabolic.

The same logic used in solving the collision integral for electron-electron collisions (in Appendix \ref{int_eqn_reduction}) can be applied here, with the modification
\begin{equation} \label{ak_trans_holes}
	d\kv'\, d\pv\, = \frac{m_c m_v^2}{2\hbar^6\cos\l(\theta/2\r)} d\ve_{k'}\, d\ve_p\, d\ve_{p'}\, \sin\theta d\theta\, d\varphi\, d\varphi_p\, ,
\end{equation}
where $m_v$ is the effective valence band hole mass. Following the same procedure used in \ee scattering, the average effective scattering time for electron-hole collisions is
\begin{equation} \label{scat_avg_eh}
	\tau^{*(h)}_{\text{avg}} = \frac{B^{(h)}}{2\pi^2} \sum_{l=0}^{\infty} \frac{4l+3}{\Lambda_{2l}\l[\Lambda_{2l}-\lambda_1\r]} \, ,
\end{equation}
where the defines are the same as in Eqs.~\eqref{BFull}-\eqref{lambda2} and
\begin{equation}
	B^{(h)} = \frac{\hbar^7(2\pi)^4}{m_c m_v^2 \l(k_BT\r)^2} A_1^{-1} \label{Bh} \, .
\end{equation}

The plane wave scattering time is simply
\begin{equation}
	\tau^{(h)}_k = 2B^{(h)} \l[\l(\frac{\ve_k-\mu}{k_BT}\r)^2+\pi^2\r]^{-1} \, 
\end{equation}
with an average of
\begin{equation}
	\tau^{(h)}_{\text{avg}} = \frac{3B^{(h)}}{2\pi^2} \, .
\end{equation}

\subsection{Electron-impurity collisions}
The collision integral for elastic collisions from impurities is considerably simpler:
\begin{equation} \label{coll_int_imp}
	I_{\kv}(t)^{\text{e-i}} = -\frac{\pi}{\hbar} \sum_{\kv'} \l| W_{\kv\kv'} \r|^2 \delta\l(\ve_k-\ve_{k'}\r) \l(\rho_{1\kv}-\rho_{1\kv'}\r) \, .
\end{equation}
The resulting effective scattering rate is\cite{fabian_aps,opt_orient}
\begin{equation} \label{final_scat_imp}
	\frac{1}{\tau^{*(i)}_{k_F}} = \frac{m_c k_F n_i}{4\pi\hbar^3} \int_{-1}^1 d(\cos\theta)\, \l|W(\cos\theta)\r|^2 \l[1-P_3(\cos\theta)\r] \, ,
\end{equation}
where $n_i$ is the density of impurities.

\section{Effective scattering amplitudes} \label{scat}
We can apply the formulas derived in the previous section to the calculation of effective scattering times in both intrinsic and extrinsic semiconductors, provided they are in the degenerate regime ($k_BT\ll \ve_F$). The intrinsic case will be considered first, since it is in this case that many-body effects play the dominant role. Electron-hole pairs are created in an intrinsic semiconductor by optical excitation and may, if the recombination time is sufficiently long, condense in a degenerate electron-hole liquid. The impurity concentration is negligible and electron-electron and, especially, electron-hole scattering plays the dominant role in controlling the DP mechanism of spin relaxation. We will also consider, for completeness, the electron liquid in extrinsic $n$-type doped semiconductors. In this case the doped electrons come from donor impurities and there is typically one impurity per electron. Furthermore, because we are considering a bulk semiconductor, the impurities are homogeneously distributed throughout the electron liquid, and thus provide the dominant scattering mechanism -- far more important than electron-electron scattering, as we will see.

The crucial ingredient of the calculation is, of course, the effective interaction $W$ to be used in Eqs.~\eqref{scat_avg_ee}, \eqref{scat_avg_eh}, and \eqref{final_scat_imp}. Here we have two options. The first option is to adopt a simple RPA-screened Coulomb interaction (also known as Lindhard screening); this leads to a parameter-free expression for the interaction, which should be exact in the high density limit. However the Lindhard screening is known to be inaccurate as the density decreases, and furthermore the form of the RPA effective interaction misses important correlations between the electrons under consideration and the surrounding medium -- correlations that become more and more important at low density. To counter these drawbacks one may resort to a second option, in which both the dielectric function and the effective interaction are modified by many-body local field factors which encapsulate exchange and correlation effects. (Discussion of these modified effective interactions can be found in Section 5.5 of reference [\onlinecite{vignale_book}].) Unfortunately, even this approach is not problem-free, since the local field factors are imperfectly known in the electron liquid, and even more so in the electron-hole liquid. Nevertheless, the behavior of the many-body local field factors is constrained by exact sum rules and limiting cases which, taken together, allow us to form a qualitatively correct picture of the effective interaction at low density (still in the degenerate regime). We would like to point out that J. Zhou has also performed calculations of the spin relaxation time including Singwi-Tosi-Land-Sj\"{o}lander local field corrections, but in 2D GaAs systems.\cite{Zhou200850}

We define the two-particle bare Coulomb interaction ($v_{ij}$) and static Lindhard function ($\chi_{0i}$) ahead of time:
\begin{align}
	v_{ij}(q) &= \frac{4\pi e_i e_j}{\epsilon_r q^2} \, ,\\
	\chi_{0i}(q) &= -N_i(0) \l[\frac{1}{2} + \frac{1-x^2}{4x} \ln \l|\frac{1+x}{1-x}\r|\r] \, ,
\end{align}
where $x=q/2k_F$, $N_i(0)$ is the density of states at the Fermi level for electrons or holes, and $\epsilon_r$ is the relative dielectric constant of the medium.

For the electron liquid, electron-electron interactions are handled by a spin-averaged Kukkonen-Overhauser scattering amplitude:\cite{prb_20_550}
\begin{equation} \label{KO}
	W_{e-e}^{\text{EL}}(q) = v_{11}(q) + \l[ v_{11}(q)\l(1-G_{11}^s(q)\r) \r]^2\chi_{11}(q)\, ,
\end{equation}
with the static density-density response function:
\begin{equation} \label{chi_nn}
	\chi_{11}(q) = \frac{\chi_{01}(q)}{1-v_{11}(q)\l[1-G_{11}^s(q)\r]\chi_{01}(q)} \, .
\end{equation}
$G_{11}^s(q)$ is the spin symmetric local field factor for electron liquids. The scattering amplitude has the following physical interpretation: The first term $v_{11}(q)$ is the bare Coulomb interaction between electrons while the second term has an electron interacting with the density induced in the electron liquid which in turn acts on another electron (each with a reduced Coulomb interaction). Our calculations in the following section use a fit for $G_{11}^s(q)$ from Moroni, et. al.\cite{prl_75_689} (also reviewed in [\onlinecite{prb_57_14569}]) which is based on diffusive Monte Carlo studies of the static density-density response function. It should be noted that $G_{11}^s(q)$ used in these calculations is intended for use with $2<r_s<10$ and $0<q<3k_F$, but limiting behavior and graphical analysis suggests the fit is still qualitatively reasonable for the range of $r_s$ examined in this paper, $0.4<r_s<2$. Since we are restricted to the Fermi surface, the range of $q$ is not a concern, $0<q<2k_F$. Setting $G_{11}^s(q)=0$ amounts to using the random phase approximation (RPA).

Impurity scattering rates in the electron liquid are calculated with the electron-test charge effective scattering amplitude:
\begin{equation}
	W^{\text{EL}}_{e-i}(q) = v_{21}(q) + v_{21}(q)\chi_{11}(q)v_{11}(q)\l[1-G_{11}^s(q)\r] \, ,
\end{equation}
where $\chi_{11}(q)$ is as in Eq.~\eqref{chi_nn} and we are assuming a uniform distribution of impurities. The physical interpretation is similar to that described for Eq.~\eqref{KO}, except now the impurity interacts with the density induced in the electron liquid with the bare Coulomb interaction.

For the electron-hole liquid, a multicomponent scattering amplitude analogous to the Kukkonen-Overhauser formula is necessary:
\begin{multline}
	W_{ij}^{\text{EHL}}(q) = v_{ij}(q) \\
	+ \sum_{kl=1}^2 v_{ik}(q)\l[1-G^s_{ik}(q)\r] \chi_{kl}(q)v_{lj}(q)\l[1-G^s_{lj}(q)\r] \, .
\end{multline}
Electron-electron scattering is represented by $W_{11}^{\text{EHL}}$, and electron-hole scattering by $W_{12}^{\text{EHL}}$. The density response function is found from\cite{prb_31_2729}
\begin{equation}
	[(\utilde{\chi})^{-1}]_{ij}(q) = \l[\chi_{0i}(q)\r]^{-1}\delta_{i,j} - v_{ij}(q)\l(1-G^s_{ij}(q)\r) \, ,
\end{equation}
where $\utilde{\chi}$ is the spin symmetric density response function in matrix form. The local field factors for electron-hole liquids (which should not be confused to be the same as those for the electron liquid) have not been well studied to the best of our knowledge. We opt to use an approximate form of $G^s_{ij}(q)$ from [\onlinecite{prb_31_2729}] with parameters based on a hole-to-electron mass ratio of $m_h/m_e=6$, which is actually not too far from the value of the ratio in GaAs. Again, setting $G^s_{ij}(q)=0$ amounts to using RPA.

\section{Calculations of the effective scattering rates and spin relaxation times} \label{disc}

According to Eq.~(\ref{tau_fin-degen}) to calculate the spin relaxation time $\tau^{(s)}$ we need two ingredients: the average scattering time $\tau^*_{\text{avg}}$ and the mean square of the spin-orbit field $\langle{\Omega_{\kv}^2}\rangle_{k_F}$. The former has been calculated in Sec.~III and is given by Eq.~(\ref{scat_avg_ee}). The later can be found straightforwardly using the explicit form of $\Omega_{\kv}$ in Eq.~(\ref{Omega}):
\begin{equation}
	\frac{2}{3}\langle{\Omega_{\kv}^2}\rangle_{k_F} = \frac{32}{105} \frac{\alpha_c^2 \ve_F^3}{\hbar^2 E_g} \, .
\end{equation}
Inserting this result into Eq.~(\ref{tau_fin-degen}) we get for the spin relaxation time
\begin{equation}
	\frac{1}{\tau^{(s)}} = \tau^*_{\text{avg}} \frac{32}{105} \frac{\alpha_c^2 \ve_F^3}{\hbar^2 E_g} \, ,
\end{equation}

\begin{figure}[thb]
\begin{center}
	\includegraphics[width=8cm]{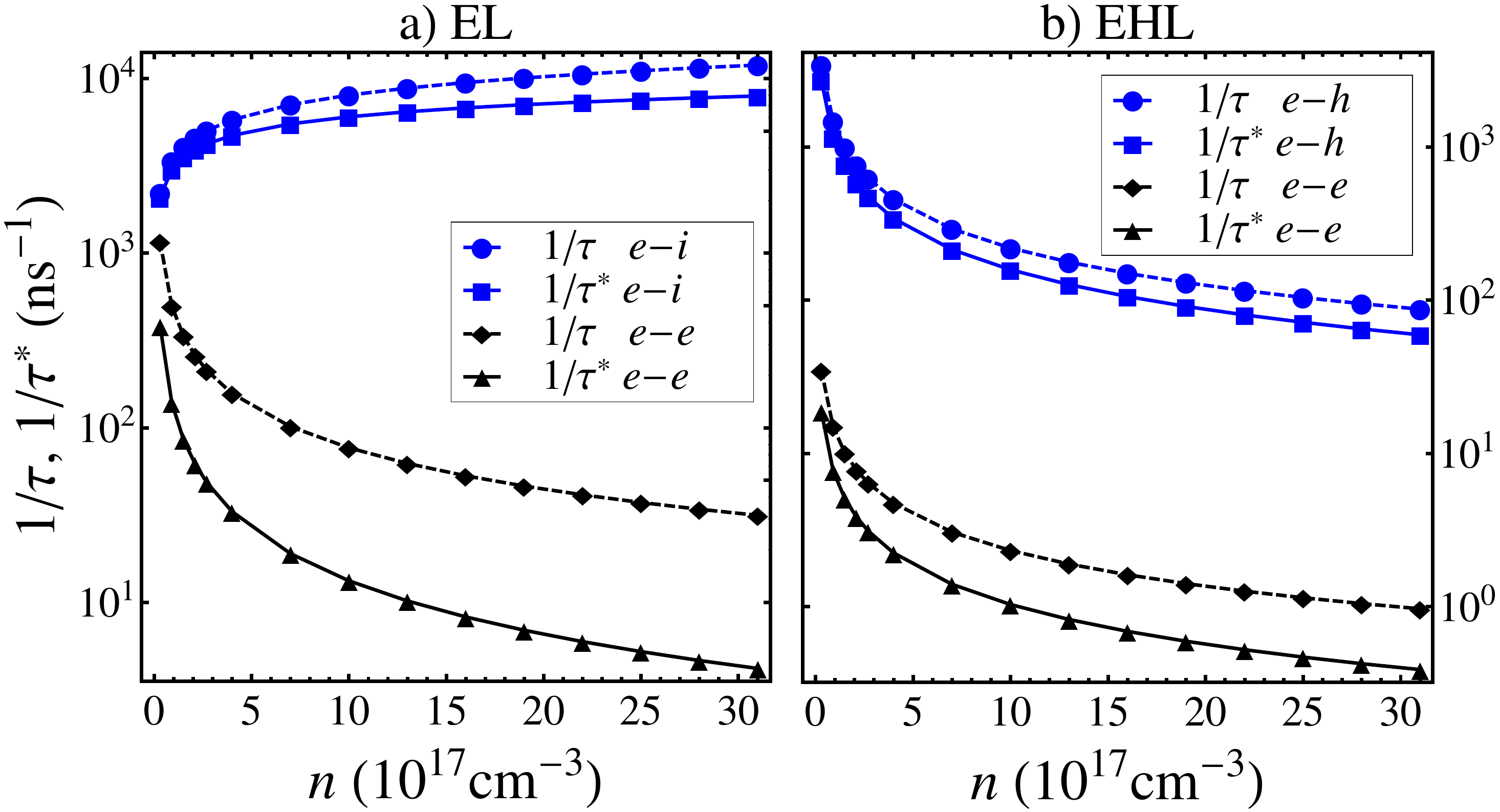}
\end{center}
\caption[effective scattering rates]{(Color online) The plane wave and effective scattering rates in the a) electron liquid (EL) and b) electron-hole liquid (EHL) have been calculated with RPA at $T=20$K.}
\label{fig:GaAsESRvsMOMrpa}
\end{figure}

While the above results are valid for generic zincblende III-V semiconductors, we opt to include results for GaAs with the following properties: $m_c = 0.067m_e$, $m_v = 0.47m_e$, $E_g = 1.43$eV, $\epsilon_r=13.2$, and $\alpha_c = 0.07$. In Fig.~\ref{fig:GaAsESRvsMOMrpa}, panel (a), the electron scattering rates from electrons and impurities in the electron liquid (EL) are plotted as functions of density. Panel (b) shows scattering from electrons and holes in the electron-hole liquid (EHL). Recalling that the SRT is proportional to the scattering rate, we can see immediately via application of Matthiesen's rule that between \ee and \ei, the SRT will be controlled by \ei collisions in $n$-GaAs. Only at low densities do \ee collisions begin to make a significant contribution, and at that point degeneracy diminishes so that one should be more careful about including non-degenerate effects. In the electron-hole liquid (intrinsic GaAs), \ee collisions are downplayed even more and \eh collisions dominate the effective scattering rate. The physical reason for this is that holes, with their large mass and concurrently high density of states provide effective screening of the electron-electron interaction, which thus turns out to be much smaller than in the electron liquid. On the other hand, the presence of electron-hole scattering more than makes up for what is lost in the electron-electron scattering strength.

When compared to the scattering rates of plane waves, the effective scattering rates are found to be generally smaller: see, again, Fig.~\ref{fig:GaAsESRvsMOMrpa}. Approximating the effective scattering time by the plane wave lifetime can lead to a difference in the SRT of between $0$ and $1$ order of magnitude. As discussed in the introduction, this happens because some collision processes cause a large change in momentum but a small change in the Dresselhaus field (consider for example a collision that takes us from a point in k-space where the Dresselhaus field vanishes to another symmetry-related point where it also vanishes); such a process contributes to the plane wave lifetime, but has no effect on spin relaxation. The use of the effective scattering rate rather than the plane wave approximation appears to be of utmost importance for high densities, but one must be cautious that the rate of spin relaxation does not approach the order of magnitude of the momentum scattering rate, as we will see momentarily.

\begin{figure}[thb]
\begin{center}
	\includegraphics[width=8cm]{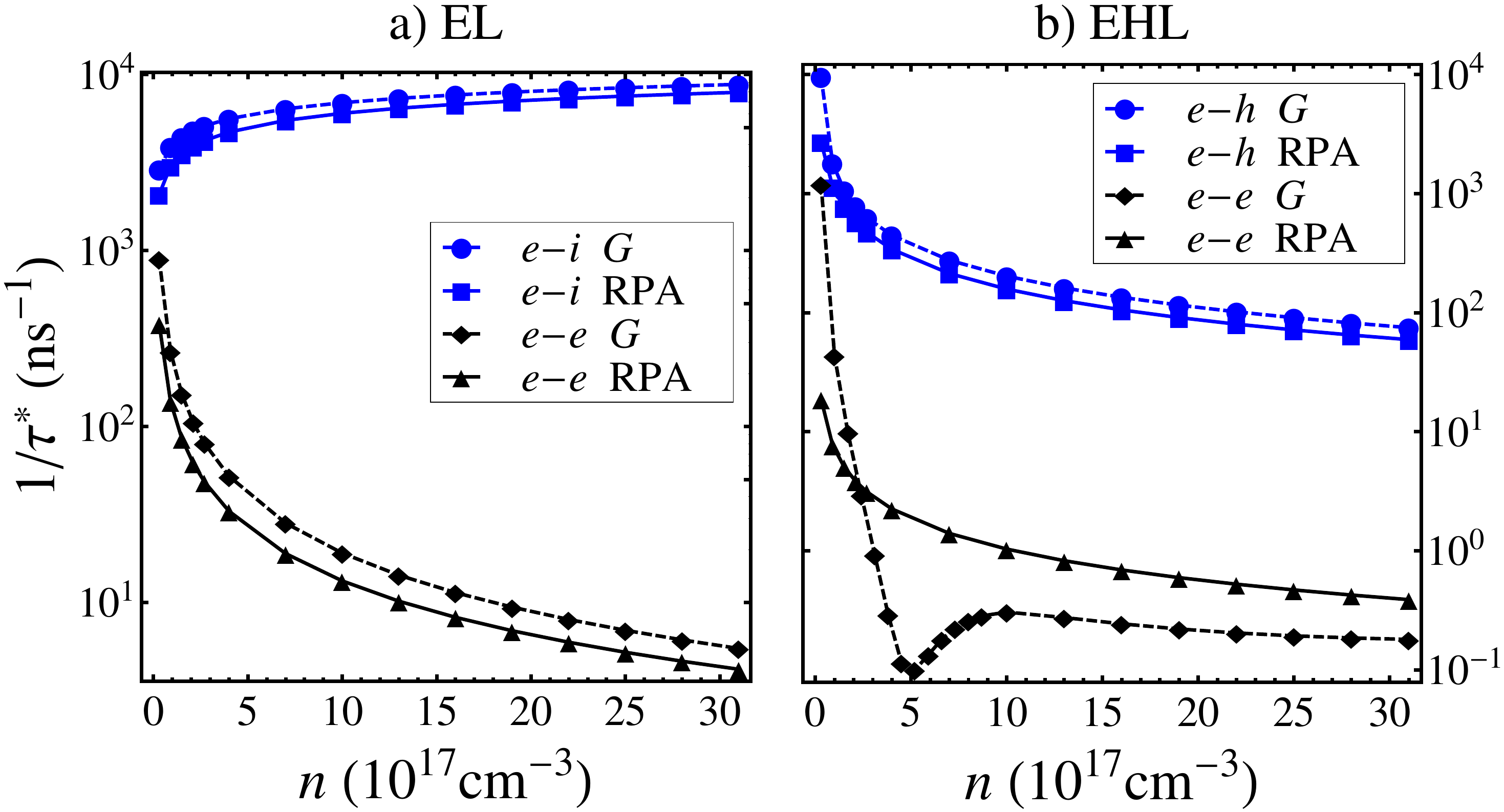}
\end{center}
\caption[local field factor contributions]{(Color online) a) Local field factors for the electron liquid (EL) are derived from fits by Moroni, \textit{et al.}\cite{prl_75_689} and their inclusion in the scattering amplitude has the overall effect of enhancing the scattering rate. b) Local field factors for the electron-hole liquid (EHL) are based on a hole-to-electron mass ratio of 6 and are found in Vignale, \textit{et al.}\cite{prb_31_2729}. All rates have been calculated at $T=20$K.}
\label{fig:GaAsGvsRPAesr}
\end{figure}

In Fig.~\ref{fig:GaAsGvsRPAesr} the effect of local field factors on the effective scattering rates is apparent especially at low densities. While the local field factors for the electron-hole liquid are rough approximations based on $m_c/m_v=6$, and so their results might not be quantitatively accurate, they likely reflect the qualitative trends as functions of density. Additionally, while the local field factors for the electron liquid have been calculated using modern quantum Monte Carlo analyses of the density response function, they were designed for the pure electron liquid (no impurities) and considerably smaller densities. Disclaimers aside, we see that the inclusion of the local field factor generally enhances the effective scattering rates, and therefore suppresses the spin relaxation rate (\ee scattering in the EHL is a bit of an anomaly here). The enhancement of the interaction is expected on physical grounds since, as the density is lowered, the electron-hole liquid becomes increasingly ``soft" -- meaning a large density of low-lying excitations -- and has a strong tendency to develop inhomogeneous density waves. The strong effective interaction arising from this softening was recognized long ago as a possible mechanism of superconductivity in the electron-hole liquid.\cite{prb_31_2729} However, the precise value of $r_s$ (the average inter-particle distance in units of the effective Bohr radius) at which the transition would occur, as well as the actual degree of enhancement at a given $r_s$, are still quite uncertain.

With regard to Zhao's work mentioned in the introduction, the claims argued there are still relevant. The spin polarized packets in Zhao's experiment consist of a degenerate center and non-degenerate tails. In the tails, \ei collisions are reduced in liu of \ee collisions so that the effective scattering rate is generally smaller than in the center of the packet. The net effect is that spins relax faster in the tails than in the center.

\begin{figure}[thb]
\begin{center}
	\includegraphics[width=8cm]{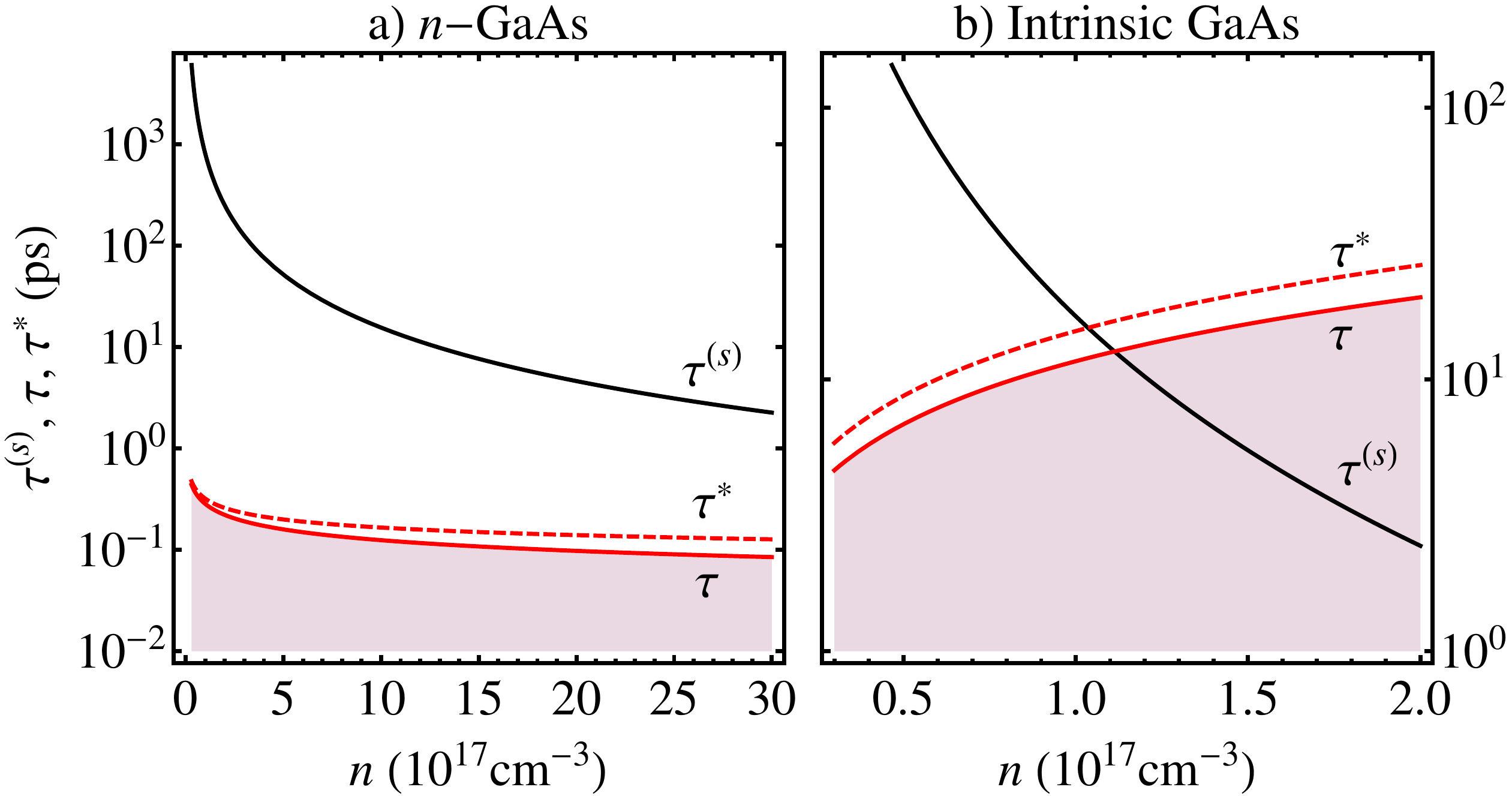}
\end{center}
\caption[GaAs SRT]{(Color online) The spin relaxation time in a) $n$-GaAs and b) intrinsic GaAs. Calculations were performed with RPA at $T=20$K. $n$-GaAs is modeled by the EL and includes contributions from electron-electron and electron-impurity collisions. Impurity collisions dominate the SRT for the degenerate electron liquid. Intrinsic GaAs is modeled by the EHL and includes contributions from electron-electron and electron-hole collisions. Electron-hole collisions dominate the SRT for the degenerate electron-hole liquid. Where the SRT crosses the scattering time marks the beginning of the breakdown of the standard DP assumptions. From this point on, spin relaxation and momentum relaxation occur on similar time scale and can no longer be separated. Eventually, at very high density, momentum relaxation becomes extremely slow and spin and momentum dynamics become decoupled again, this time with each spin performing an independent precession in the Dresselhaus field at a given point $\kv$.} 
\label{fig:GaAsSRT}
\end{figure}

The spin relaxation times for $n$-GaAs and intrinsic GaAs are plotted in Fig.~\ref{fig:GaAsSRT}. As $n$-GaAs is largely dominated by \ei collisions, it can be understood why past theoretical curves fit the experimental data in references [\onlinecite{physica_e_10_1_1}] and [\onlinecite{prb_66_245204}] so well. Caution should be taken when calculating the SRT for electrons in the electron-hole liquid. A region of validity becomes apparent for relaxation time approximation in Eq.~\eqref{r1k_relation}. As pointed out earlier, SRTs shorter than the plane wave scattering time break the derivation of the effective scattering time. The plane wave lifetime in $n$-GaAs is sufficiently short to avoid this problem for all densities examined here, but \eh scattering has a comparatively long scattering lifetime which results in unfeasibly short SRTs. Admittedly, the overall effective scattering rate ought to take into account all scatterers (phonons, impurities, etc.) which may ultimately lengthen the SRT. The intrinsic GaAs example here is an idealized electron-hole liquid.

\section{Conclusion} \label{conc}

We have derived simple one- and two-dimensional integrals for the effective scattering rate of electrons in the many body system, valid in the degenerate regime. The energy dependence has been handled according to exact expressions by Sykes/Brooker\cite{SykesBrooker} and the angular dependence is a result of a Dresselhaus field acting as an addition spin-axis re-orientation mechanism. In general, scattering events with the Dresselhaus field present are less effective in randomizing the momentum axis, than without a field present. 

In highly degenerate systems, the contribution of electron-electron scattering to the spin relaxation time is minimal in comparison to electron-impurity or electron-hole contributions. For the case of high degeneracy in intrinsic GaAs, we observe a limitation of the assumption that the timescale of spin relaxation is long compared to the scattering lifetime. In this regime the quasiparticles are essentially non-interacting and the DP mechanism is superseded by the spin precession of individual quasiparticles of essentially constant momentum.

Local field factors, taking into account exchange and correlation effects, have been introduced into the scattering rate calculations. Scattering rates are generally enhanced compared to using RPA, more-so at low densities, for both electron-electron interactions and electron-hole interactions. Significant improvements to the accuracy of the scattering rates could be made with new local field factors tailored to the densities of typical intrinsic and $n$-type III-V semiconductors.\\

\section{Acknowledgments}
This work was supported by the National Science Foundation under grant number DMR-0705460 and the Department of Energy under grant number DE-FG02-05ER46203. The work of IVT was supported by Spanish MEC (FIS2007-65702-C02-01), ``Grupos Consolidados UPV/EHU del Gobierno Vasco'' (IT-319-07), and the European Union through e-I3 ETSF project (Contract No. 211956). We are indebted to Ming-Wei Wu for a crucial discussion on the relevancy of many-body effects in the photo-excited electron-hole liquid in intrinsic semiconductors.

\appendix

\section{Equivalence of harmonics in Eq.~(\ref{coll_int_reduced}) and the relaxation time approximation} \label{r1x_to_omega_k}
It is relatively straightforward to demonstrate that integration of $\Omega_{\qv,z}$ (where $\qv$ can be any of $\kv'$, $\pv$, or $\pv'$) over $d\qh$ in Eq.~(\ref{coll_int_reduced}) results in a term proportional to $\Omega_{\kv,z}$. Let us write $\Ov_{\qv,z}$ in terms of spherical harmonics:
\begin{align}
	\Omega_{\qv,z} = \Omega_0 q^3 \sqrt{\frac{8\pi}{105}} \l[Y_3^2(\vt_q,\vp_q) - Y_3^{-2}(\vt_q,\vp_q)\r]
\end{align}
where $\Omega_0$ is a constant.

The angular portion of the collision integral in Eq.~\eqref{coll_int_reduced} has each $\Omega_{\qv,z}$ integrated with a scattering probability. These scattering probabilities can be expanded in Legendre polynomials of argument $\cos\alpha = \kh\cdot\qh$:
\begin{equation}
	W = \sum_l w_l P_l\l(\cos\alpha\r) \, ,
\end{equation}
where $W$ can represent either $\l|W_{\kv \pv \kv' \pv'}\r|^2$ or $\l|W_{\kv \pv \kv' \pv'}W_{\kv \pv \pv' \kv'}\r|$. The coefficients can later be found with
\begin{equation}
	w_l = \frac{2l+1}{2} \int_{-1}^1 d\l(\cos\alpha\r) W P_l\l(\cos\alpha\r) \, .
\end{equation}
The addition theorem for spherical harmonics expands $P_l$ in spherical harmonics:
\begin{equation}
	P_l\l(\cos\alpha\r) = \frac{4\pi}{2l+1} \sum_m Y_l^{*m}\l(\vt_q,\vp_q\r) Y_l^m\l(\vt_k,\vp_k\r) \, .
\end{equation}
Then, integration over $d\qh$ is easy due to the orthogonality of spherical harmonics:
\begin{align}
	\int d\qh\, W \Omega_{\qv,z} =& w_3 \frac{4\pi}{7} \Omega_{\kv,z} \nonumber \\
	=& 2\pi \Omega_{\kv,z} \int_{-1}^1 d(\cos\alpha)\, W P_3(\cos\alpha) \, .
\end{align}
That $\Omega_{\kv,z}$ can be extracted from each term in Eq.~\eqref{coll_int_reduced} goes to show that the relaxation time approximation in Eq.~\eqref{RTA} and an angular-independent $\tau^*_k$ are both reasonable claims.

\section{Reduction of the integral equation for the effective scattering time in electron-electron collisions} \label{int_eqn_reduction}

\begin{figure}[htb]
\begin{center}
	\includegraphics[width=8cm]{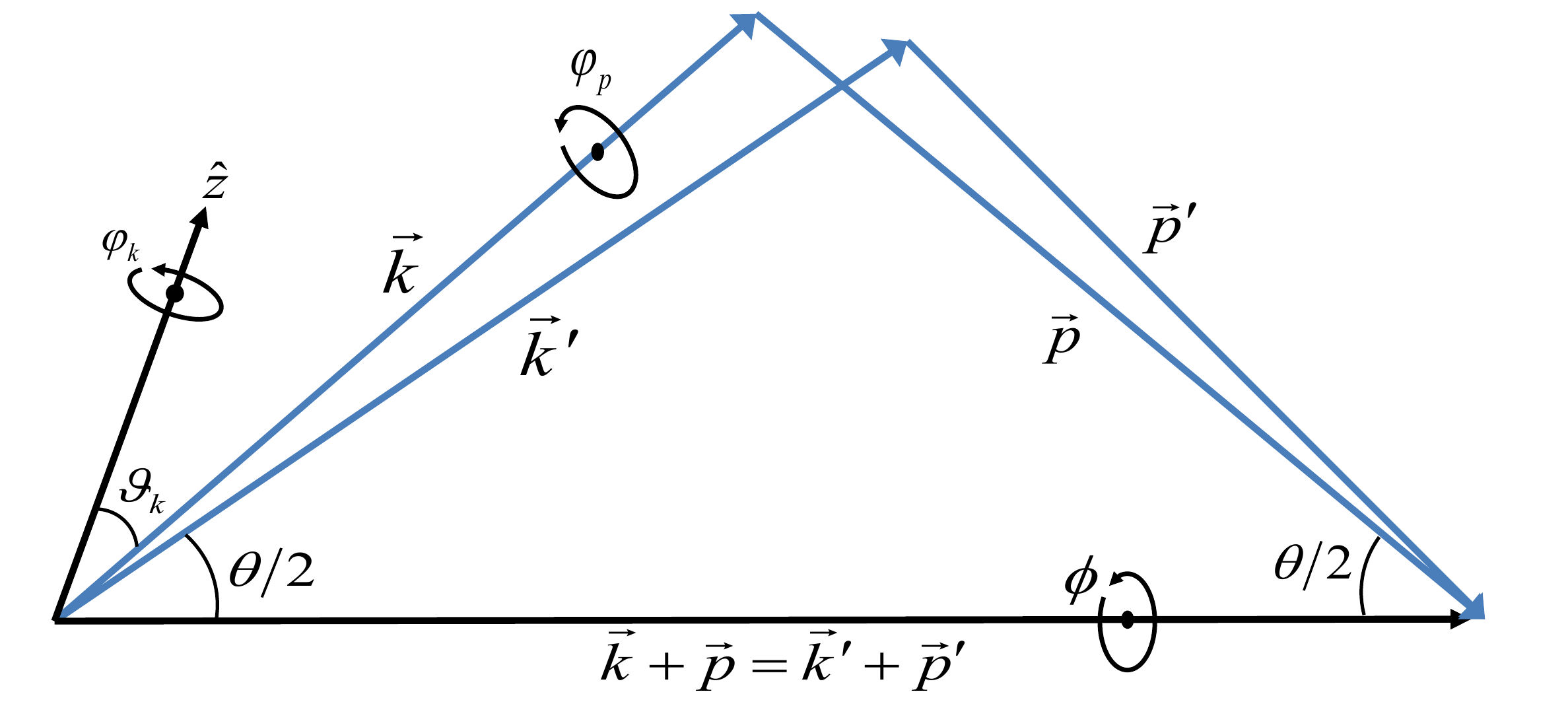}
\end{center}
\caption[AK angles]{(Color online) For an arbitrary direction of $\kv$ with respect to $\zh$, the direction of each momentum vector can be put entirely in terms of $\kv$ and the angles $\theta$, $\phi$, and $\vp_p$. Angle $\phi$ defines the planes in which $\kv$,$\pv$ and $\kv'$,$\pv'$ lie.}
\label{fig:ak_angles}
\end{figure}

We begin our solution for $\tau^*_k$ by working from Eq.~\eqref{coll_int_reduced} with the replacements discussed, cubic symmetry allows $\Ov_{\kv}\times\sh \rightarrow \Omega_{\kv,z}$ and angular independence in $\tau^*_{\kv}$ allows $\tau^*_{\kv} \rightarrow \tau^*_k$:
\begin{widetext}
\begin{multline} \label{coll_int_reduced2}
	f^0_k (1-f^0_k)\Omega_{\kv,z} = \frac{4\pi}{\hbar} \frac{1}{\l(2\pi\r)^6} \sum_{\pv'} \int d\kv'\, d\pv\, f^0_k(1-f^0_{k'})f^0_p(1-f^0_{p'}) \delta\l(\ve_k+\ve_p-\ve_{k'}-\ve_{p'}\r) \delta_{\kv+\pv,\kv'+\pv'} \\
	\l\{ \l|W_{\kv \pv \kv' \pv'}\r|^2 \l[\tau^*_{k} \Omega_{\kv,z} - \tau^*_{k'} \Omega_{\kv',z} \r] - \frac{1}{2}\l|W_{\kv \pv \kv' \pv'}W_{\kv \pv \pv' \kv'}\r| \l[\tau^*_{k} \Omega_{\kv,z} - \tau^*_{k'} \Omega_{\kv',z} + \tau^*_{p} \Omega_{\pv,z} - \tau^*_{p'} \Omega_{\pv',z}\r] \r\} \, .
\end{multline}
\end{widetext}
The sum over $\pv'$ can be evaluated immediately using conservation of momentum. Then, following a method introduced by Abrikosov and Khalatnikov\cite{abrikosov_tfl} (or for an alternative derivation, Baym and Pethick\cite{baym_pethick_lflt}), at low temperatures the integral over momentum space can be re-expressed as
\begin{equation} \label{ak_trans}
	d\kv'\, d\pv\, = \frac{m_c^3}{2\hbar^6\cos\l(\theta/2\r)} d\ve_{k'}\, d\ve_p\, d\ve_{p'}\, \sin\theta d\theta\, d\phi\, d\varphi_p\, ,
\end{equation}
where $\theta$ is the angle between $\kv$ and $\pv$, and $\phi$ is a polar angle about the $\kv+\pv$ axis (See Fig.~\ref{fig:ak_angles}). Aside from spin dependence, the scattering amplitudes are functions only of $\theta$ and $\phi$.

At this point our kinetic equation has the form 
\begin{widetext}
\begin{multline} \label{intermediate}
	\Omega_{\kv,z} = \frac{2m_c^3}{\hbar^7(2\pi)^5} \int_{-\infty}^{\infty} d\ve_{k'} \int_{-\infty}^{\infty} d\ve_{p} \int_{-\infty}^{\infty} d\ve_{p'}\, \l(\frac{1-f^0_{k'}}{1-f^0_k}\r)f^0_p(1-f^0_{p'}) \delta\l(\ve_k+\ve_p-\ve_{k'}-\ve_{p'}\r) \int_0^{2\pi} d\phi \int_0^{\pi} d\theta \int_0^{2\pi} d\vp_p \\
	\sin(\theta/2) \l\{ \l|W_{\kv \pv \kv' \pv'}\r|^2 \l[\Omega_{\kv,z}\tau^*_{k} - \Omega_{\kv',z}\tau^*_{k'}\r] - \frac{1}{2}\l|W_{\kv \pv \kv' \pv'}W_{\kv \pv \pv' \kv'}\r| \l[\Omega_{\kv,z}\tau^*_{k} - \Omega_{\kv',z}\tau^*_{k'} + \Omega_{\pv,z}\tau^*_{p} - \Omega_{\pv',z}\tau^*_{p'}\r] \r\} \, .
\end{multline}
\end{widetext}
The spherical harmonic definition of $\Omega_{\kv,z}$ from Eq.~\eqref{omega_z} is useful at this point. The orthogonality associated with the spherical harmonic functions is exploited by multiplying through by $\Omega_{\kv,z}$ in Eq.~\eqref{intermediate} and integrating over $\sin\vt_k d\vt_k\, d\vp_k$. The left hand side reduces with use of
\begin{equation}
	\int_0^{\pi} d\vt_k\, \int_0^{2\pi} d\vp_k\, \sin\vt_k \Omega_{\kv,z}^2 = C^2 k^6 \frac{16\pi}{105} \, ,
\end{equation}
where $C^2=\alpha_c^2 \hbar^4 / (2m_c^3E_g)$.

There is still the matter of putting the angles corresponding to vectors $\pv$, $\kv'$, and $\pv'$ in in terms of the integration variables. This is accomplished by introducing a right hand orthonormal basis of vectors:
\begin{subequations}
\begin{align}
	\label{khat} \kh &= \l\{\sin\vt_k\cos\vp_k,\sin\vt_k\sin\vp_k,\cos\vt_k\r\}\, , \\
	\th &= \l\{\cos\vt_k\cos\vp_k,\cos\vt_k\sin\vp_k,-\sin\vt_k\r\}\, , \\
	\nh &= \kh \times \th \, .
\end{align}
\end{subequations}
Aligning $\kh$ along $\zh$, it is seen that
\begin{equation} \label{phat}
	\ph = \cos\theta \kh + \sin\theta\cos\vp_p \th + \sin\theta\sin\vp_p \nh \, .
\end{equation}
A rotation formula can be applied so that vectors $\kv$ and $\pv$ are turned about the $\kv+\pv$ axis to find $\kh'$ and $\ph'$:
\begin{align}
	\kh' =& \frac{\kh+\ph}{2} + \frac{\kh-\ph}{2}\cos\phi + \l(\frac{\kh\times\ph}{|\kh+\ph|}\r)\sin\phi\, , \\
	\ph' =& \frac{\kh+\ph}{2} - \frac{\kh-\ph}{2}\cos\phi - \l(\frac{\kh\times\ph}{|\kh+\ph|}\r)\sin\phi\, .
\end{align}
Standard relations are used to put the angular components of $\kh'$, $\ph$, and $\ph'$ in terms of the integration variables:
\begin{align}
	\label{ang_from_vec_z} \cos \vt_q &= q_z \, , \\
	\label{ang_from_vec_x} \cos\vp_q &= \frac{q_x}{\sqrt{1-q_z^2}} \, .
\end{align}

This concludes the needed angular transformations. Let us define the angular-only portion of $\Omega_{\kv,z}$ with $\kappa(\vt_k,\vp_k)$. In other words
\begin{align} \label{kappa}
	\kappa(\vt_k,\vp_k) =& \cos\vartheta_k \sin^2\vartheta_k \cos(2\varphi_k) \nonumber \\
	=& (\cos\vt_k-\cos^3\vt_k)(2\cos^2\vp_k-1) \, ,
\end{align}
where identities have been used to write $\kappa$ in terms of cosine functions for direct use of Eqs.~\eqref{ang_from_vec_z} and \eqref{ang_from_vec_x}. This leaves us with

\begin{widetext}
\begin{multline} \label{explicit_int}
	\frac{16\pi}{105} = \frac{2m_c^3}{\hbar^7(2\pi)^5} \int_{-\infty}^{\infty} d\ve_{k'} \int_{-\infty}^{\infty} d\ve_{p} \int_{-\infty}^{\infty} d\ve_{p'}\, \l(\frac{1-f^0_{k'}}{1-f^0_k}\r)f^0_p(1-f^0_{p'}) \delta\l(\ve_k+\ve_p-\ve_{k'}-\ve_{p'}\r) \\
	\int_0^{2\pi} d\vp_k \int_0^{\pi} d\vt_k \int_0^{2\pi} d\phi \int_0^{\pi} d\theta \int_0^{2\pi} d\vp_p \sin\vt_k \sin(\theta/2) \kappa\l(\vt_k,\vp_k\r) \l\{ \l|W_{\kv\kv'}\r|^2 \l[\kappa\l(\vt_k,\vp_k\r)\tau^*_{k} - \frac{\kappa\l(\vt_{k'},\vp_{k'}\r)}{(k/k')^3}\tau^*_{k'} \r] \r.\\
	 - \l. \frac{1}{2}\l|W_{\kv\kv'}W_{\kv\pv'}\r| \l[\kappa\l(\vt_k,\vp_k\r)\tau^*_{k} - \frac{\kappa\l(\vt_{k'},\vp_{k'}\r)}{(k/k')^3}\tau^*_{k'} + \frac{\kappa\l(\vt_p,\vp_p\r)}{(k/p)^3}\tau^*_{p} - \frac{\kappa\l(\vt_{p'},\vp_{p'}\r)}{(k/p')^3}\tau^*_{p'} \r] \r\} \, .
\end{multline}
\end{widetext}
Here, we have made use of the fact that the scattering amplitude in the direct portion of the collision integral only depends on the momentum transfer $\kv\rightarrow\kv'$, whereas the exchange portion includes the possibilities of $\kv\rightarrow\kv'$ and $\kv\rightarrow\pv'$. For scattering near the Fermi surface, the scattering amplitudes only vary with the angle between incoming and outgoing momenta. These angles can be written in terms of just $\theta$ and $\phi$ with the following relations:
\begin{align}
	\begin{split}
	\cos\theta_1 &= \kh\cdot\kh' \\
	&= \frac{1}{2} (1 + \cos\theta + \cos\phi - \cos\theta\cos\phi) ,
	\end{split} \\
	\begin{split}
	\cos\theta_2 &= \kh\cdot\ph' \\
	&= \frac{1}{2} (1 + \cos\theta - \cos\phi + \cos\theta\cos\phi) .
	\end{split}
\end{align}
The scattering amplitudes written as functions of $\theta_1$ and $\theta_2$ are
\begin{align}
	W_{\kv\kv'} =& W(\cos\theta_1)\, , \\
	W_{\kv\pv'} =& W(\cos\theta_2)\, .
\end{align}
For reference, the dimensionless momentum transfers are
\begin{align}
	\label{mom_trans_dir} \qt_1 &= |\kv-\kv'|/\l(2k_F\r) = \sin\l(\theta/2\r)\sin\l(\phi/2\r) \, ,\\
	\label{mom_trans_exch} \qt_2 &= |\kv-\pv'|/\l(2k_F\r) = \sin\l(\theta/2\r)\l|\cos\l(\phi/2\r)\r| \, .
\end{align}
The absolute value is a result of the range of integration, $0<\phi<2\pi$. 

Integration over $d\vp_p$, $d\vt_k$ and $d\vp_k$ can be done immediately with the result
\begin{widetext}
\begin{multline} \label{final_scat_ak}
	\frac{\hbar^7(2\pi)^4}{m_c^3} = \int_{-\infty}^{\infty} d\ve_{k'} \int_{-\infty}^{\infty} d\ve_{p} \int_{-\infty}^{\infty} d\ve_{p'}\, \l(\frac{1-f^0_{k'}}{1-f^0_k}\r)f^0_p(1-f^0_{p'}) \delta\l(\ve_k+\ve_p-\ve_{k'}-\ve_{p'}\r) \int d\Omega \\
	\l\{\frac{\l|W_{\kv\kv'}\r|^2}{\cos(\theta/2)} \l[\tau^*_{k} - \frac{P_3\l(\cos\theta_1 \r)}{(k/k')^3} \tau^*_{k'} \r] - \frac{1}{2}\frac{\l|W_{\kv\kv'}W_{\kv\pv'}\r|}{\cos(\theta/2)} \l[\tau^*_{k} - \frac{P_3\l(\cos\theta_1 \r)}{(k/k')^3} \tau^*_{k'} + \frac{P_3\l(\cos\theta \r)}{(k/p)^3} \tau^*_{p} - \frac{P_3\l(\cos\theta_2\r)}{(k/p')^3} \tau^*_{p'} \r] \r\} \, ,
\end{multline}
\end{widetext}
where $d\Omega=\sin\theta\, d\theta d\phi$ should not be confused with the Dresselhaus Larmor frequency.
 
Notice that without the Legendre polynomial terms, the scattering time could be factored out of the integral and we would have exactly the scattering rate of plane waves, as can be referenced in [\onlinecite{vignale_book}]. Given that well known solution, we are led to believe $\tau_k^*$ will have even dependences on energy and temperature.

The ratios $(k/k')^3$, $(k/p)^3$, and $(k/p')^3$ vary slowly across the Fermi surface when compared to the Fermi-Dirac functions and anticipated $\ve_k-\mu$ dependence in $\tau^*_k$. They are all set to 1. We shall consider only static screening in the scattering amplitudes, a reasonable assumption at low temperature. The resulting integral equation for $\tau^*_k$ has been exactly solved by Sykes and Brooker.\cite{SykesBrooker} We will reproduce a simplified version of the solution here.

The energy integrals are written in terms of unitless variables:
\begin{align}
	\xi &= (\ve_k - \mu)/k_BT \, ,\\
	x &= (\ve_{k'} - \mu)/k_BT \, ,\\
	y &= (\ve_{p'} - \mu)/k_BT \, ,
\end{align}
and the $\delta$-function evaluated for $(\ve_p-\mu)/k_BT$ to give
\begin{widetext}
\begin{multline}
	\frac{\hbar^7(2\pi)^4}{m_c^3 \l(k_BT\r)^2} = \int_{-\infty}^{\infty} dx \int_{-\infty}^{\infty} dy\, \l[\frac{f^0(-x)}{f^0(-\xi)}\r]f^0(x+y-\xi)f^0(-y) \int d\Omega \l\{ \frac{\l|W_{\kv\kv'}\r|^2}{\cos(\theta/2)} \l[\tau^*(\xi) - P_3\l( \cos\theta_1 \r) \tau^*(x) \r] \r. \\
	\l. - \frac{1}{2}\frac{\l|W_{\kv\kv'}W_{\kv\pv'}\r|}{\cos(\theta/2)} \l[\tau^*(\xi) - P_3\l(\cos\theta_1 \r) \tau^*(x) + P_3\l(\cos\theta \r) \tau^*(x+y-\xi) - P_3\l(\cos\theta_2\r) \tau^*(y) \r] \r\} \, .
\end{multline}
\end{widetext}
The Fermi-Dirac functions are now described by
\begin{equation}
	f^0(x) = \frac{1}{1+e^x} \, .
\end{equation}
Integration of $\tau^*(x+y-\xi)$ over $dy$ is equivalent to integration of $\tau^*(-y)$ over $dy$. This relation, along with a simple swap of $x \leftrightarrow y$ variables in terms which have $\tau^*(\pm y)$, allows the integration over $dy$ to be performed:
\begin{equation}
	\int_{-\infty}^{\infty} dy\, f^0(x+y-\xi)f^0(-y) = \frac{\xi-x}{1-e^{x-\xi}} \, .
\end{equation}
With an even dependence on $x$, we can also let $\tau^*(-x) \rightarrow \tau^*(x)$. To simplify the form of the integral equation, the angular integrals are represented by constants $B$ and $\lambda$, and the remaining $x$-dependent factor by $K(x,\xi)$:
\begin{align}
	K(x,\xi) =& \frac{f^0(-x)}{f^0(-\xi)} \l[\frac{\xi-x}{1-e^{x-\xi}}\r] \, , \\
	A_1 =& \int d\Omega\, \frac{\l|W_{\kv\kv'}\r|^2}{\cos(\theta/2)} \, , \\
	A_2 =& \int d\Omega\, \frac{\l|W_{\kv\kv'}W_{\kv\pv'}\r|}{\cos(\theta/2)} \, ,\\
	\lambda_1 =& \frac{1}{A_1} \int d\Omega\, \frac{\l|W_{\kv\kv'}\r|^2}{\cos(\theta/2)} P_3\l( \cos\theta_1 \r) \, , \\
	\begin{split}
		\lambda_2 =& \frac{1}{A_2} \int d\Omega\, \frac{\l|W_{\kv\kv'}W_{\kv\pv'}\r|}{\cos(\theta/2)} \l[P_3\l( \cos\theta_1 \r) \r. \\
		& \l. - P_3\l(\cos\theta \r) + P_3\l( \cos\theta_2 \r)\r] \, ,
	\end{split} \\
	B =& \frac{\hbar^7(2\pi)^4}{m_c^3 \l(k_BT\r)^2} \l(A_1-\frac{A_2}{2}\r)^{-1} \, ,\\
	\lambda =& \frac{A_1\lambda_1 - A_2\lambda_2/2}{A_1-A_2/2} \, ,
\end{align}
The remaining integral equation for $\tau^*(\xi)$ is exactly of the form in Sykes and Brooker:
\begin{equation}
	B = \int_{-\infty}^{\infty} dx\, K(x,\xi) \l[ \tau^*(\xi) - \lambda\tau^*(x) \r] \, .
\end{equation}
Notice that the kernel of this integral equation involves only two-dimensional angular integrals on the Fermi surface. In fact, since $A_1$ and $\lambda_1$ do not contain any exchange terms, they can actually be simplified to one-dimensional integrals of the momentum transfer $\qt_1$ in Eq.~\eqref{mom_trans_dir}:
\begin{align}
	A_1 =& 8\pi \int_0^1 d\qt_1\, \l|W(\qt_1)\r|^2 \, , \\
	\lambda_1 =& 8\pi \int_0^1 d\qt_1\, \l|W(\qt_1)\r|^2 P_3(1-2\qt_1^2) ) \, .
\end{align}

The solution for $\tau^*(\xi)$ is found by converting the integral equation to a differential equation via Fourier transform, utilizing the convolution theorem. For $\tau^*(\xi)$ with an even dependence on $\xi$,
\begin{equation}
	\tau^*(\xi) = \frac{\cosh(\xi/2)}{2\pi}\int_{-\infty}^{\infty} d\omega\, e^{-i\omega \xi} \frac{B}{\pi} \sum_{l=0}^{\infty} \frac{(4l+3)\Phi_{2l}(\omega)}{\Lambda_{2l}\l(\Lambda_{2l} - \lambda\r)} \, ,
\end{equation}
where
\begin{align}
	\Phi_l(\omega) &= p_{l+1}^1 \l(\tanh \pi \omega\r)\, , \\
	\Lambda_l &= \frac{1}{2}(l+1)(l+2)\, ,
\end{align}
and $p_l^m(x)$ are the associated Legendre polynomials.
\clearpage

\end{document}